\documentclass[authoryear,review]{elsarticle}%

\pdfoutput=1
\usepackage{amsmath, amssymb, bm, natbib}
\usepackage{graphicx}
\usepackage{srcltx}
\usepackage{rotating}
\usepackage{float}
\usepackage{lscape}
\usepackage{booktabs, subfigure}
\usepackage{verbatim, setspace}
\usepackage{calc, color}
\usepackage[english]{babel}
\usepackage{tabularx}
\usepackage{longtable}
\usepackage{multirow}
\usepackage{accents}
\usepackage{tikz}
\usepackage{colortbl}
\usepackage[none]{hyphenat}
\usepackage{ragged2e}
\usepackage{breakcites}
\usepackage{times}
\usepackage{hyperref}

\newcommand{\bmu}{\mbox{\boldmath $\mu$}}
\newcommand{\beeta}{\mbox{\boldmath $\eta$}}
\newcommand{\bphi}{\mbox{\boldmath $\phi$}}

\newcommand{\bnu}{\mbox{\boldmath $\nu$}}
\newcommand{\bSigma}{\mbox{\boldmath $\Sigma$}}
\newcommand{\bepsilon}{\mbox{\boldmath $\epsilon$}}
\newcommand{\bLambda}{\mbox{\boldmath $\Lambda$}}
\newcommand{\bbeta}{\mbox{\boldmath $\beta$}}
\newcommand{\btheta}{\mbox{\boldmath $\theta$}}
\newcommand{\bTheta}{\mbox{\boldmath $\Theta$}}

\newcommand{\bDelta}{\mbox{\boldmath $\Delta$}}
\newcommand{\bdelta}{\mbox{\boldmath $\delta$}}

\newcommand{\bzeta}{\mbox{\boldmath $\zeta$}}

\newcommand{\SN}{\textrm{SN}}
\newcommand{\ST}{\textrm{ST}}
\newcommand{\SSL}{\textrm{SSL}}
\newcommand{\SCN}{\textrm{SCN}}

\newcommand{\SMSN}{\textrm{SMSN}}
\newcommand{\A}{\mathbf{A}}
\newcommand{\B}{\mathbf{B}}

\newcommand{\bPsi}{\mbox{\boldmath $\Psi$}}

\newcommand{\btau}{\mbox{\boldmath $\tau$}}

\newcommand{\bOmega}{\mbox{\boldmath $\Omega$}}

\newcommand{\bupsilon}{\mbox{\boldmath $\upsilon$}}
\newcommand{\bUpsilon}{\mbox{\boldmath $\Upsilon$}}

\newcommand{\bGamma}{\mbox{\boldmath $\Gamma$}}

\newcommand{\be}{\mathbf{b}}

\newcommand{\yp}{\mathbf{y}}
\newcommand{\x}{\mathbf{x}}
\newcommand{\y}{\mathbf{y}}
\newcommand{\Y}{\mathbf{Y}}

\newcommand{\C}{\mathbf{C}}
\newcommand{\blambda}{\mbox{\boldmath $\lambda$}}

\newcommand{\z}{\mathbf{Z}}

\newcommand{\Z}{\mathbf{Z}}
\newcommand{\R}{\mathbf{R}}
\newcommand{\X}{\mathbf{X}}
\newcommand{\xp}{\mathbf{X}}

\newcommand{\E}{\textrm{E}}

\newcommand{\ii}{i=1,\ldots,n}

\newcommand{\balpha}{\mbox{\boldmath $\alpha$}}

\newcommand{\iid}{\stackrel {{\rm iid.}}{\sim}}
\newcommand{\ind}{\stackrel {{\rm ind.}}{\sim}}

\newcommand{\sumas}{\sum^n_{i=1}}

\newtheorem{lemma}{Lemma}
\newtheorem{corollary}{Corollary}

\newcommand{\s}{\mathbf{s}}
\newcommand{\br}{\mathbf{r}}

\newcommand{\U}{\mathbf{U}}
\newcommand{\Ha}{\textrm{H}}

\newcommand{\bi}{\textbf{b}_i}
\newcommand{\qhatk}{\widehat{Q}^{(k)}}
\newcommand{\ui}{\widehat{u}_i}
\newcommand{\taui}{\widehat{\tau}_{1i}}
\newcommand{\uki}{\widehat{u}_i^{(k)}}

\journal{Elsevier}

\begin{document}
\title{Scale mixture of skew-normal linear mixed models with within-subject serial dependence}

\author{\large Fernanda L. Schumacher$^a$, \,   Victor H. Lachos$^b$ \,  and\, Larissa A. Matos$^a$\\
{\vskip 2mm {\small{$^a$ Universidade Estadual de Campinas, Brazil}} \,\, {\small{$^b$  University of Connecticut, Storrs, 06269, USA}}}
}

\date{}

\begin{abstract}
In longitudinal studies,  repeated measures are collected over time and hence they tend to be serially correlated. In this paper we consider an extension of the skew-normal/independent linear mixed model, where the error term has a dependence structure, such as damped exponential correlation  or  autoregressive correlation of order $p$. The proposed model provides 
flexibility in capturing the effects of skewness and heavy tails simultaneously when continuous repeated
measures are serially correlated. For this robust model, we present an efficient EM-type algorithm for parameters estimation via maximum likelihood and the observed information
matrix is derived analytically to account for standard errors. The methodology is illustrated through an application to schizophrenia data and some simulation studies. The
proposed algorithm and methods are implemented in the new \textsf{R} package {\it skewlmm}.
\vspace{.2cm}
\end{abstract}


\maketitle

\section{Introduction}

Linear mixed models (LMM) are frequently used to analyze repeated measures data, because they model flexibly the within-subject correlation often present in this type of data. Usually for mathematical convenience, it is assumed that both random effect and error term follow normal  distributions (N-LMM). These restrictive assumptions, however, may result in a lack of robustness against  departures from the normal distribution and invalid statistical inferences, especially when the data
show heavy tails and skewness. For instance,  substantial bias in the maximum likelihood estimates of regression parameters can result when the random-effects distribution is misspecified \citep{drikvandi2017diagnosing,drikvandi2019nonlinear}.

To deal with this problem, some proposals have been made in the literature by replacing the assumption of normality by a more flexible class of distributions. For instance, \cite{Pinheiro01} proposed a multivariate t linear mixed model (T-LMM) and showed that it performed well in the presence of outliers. \cite{rosa2003robust} adopted a Bayesian framework to carry out posterior analysis in LMM with the thick-tailed class of normal/independent (NI-LMM) distributions. \cite{ArellanoLachos2005} proposed a skew-normal linear mixed model (SN-LMM) based on the  skew-normal (SN) distribution introduced by \cite{azzalini96}.  \cite{ho2010robust} proposed a skew-t linear mixed model (ST-LMM) based on the skew-t (ST) distribution introduced by \cite{azzalini2003distributions}. 

From a wider perspective, \cite{Lachos_Ghosh_Arellano_2009} proposed a parametric robust modeling of LMM based on skew-normal/independent (SNI) distributions, where random effects follow a SNI distribution and within-subject errors follow a NI distribution, so that observed responses follow a SNI distribution, and they define what they call the skew-normal/independent linear mixed model (SNI-LMM). They presented an efficient EM-type
algorithm for the computation of maximum likelihood (ML) estimates of parameters on the basis of the hierarchical formulation of the SNI class. It is important to note that the SNI class is a subclass of the scale mixture of skew-normal (SMSN) class introduced by \cite{Branco_Dey01}, which will be considered in this paper. More recently, \cite{pereira2019nonlinear} developed asymmetric nonlinear regression models with mixed-effects by assuming that the random components of the model follow distributions from the SMSN class. From a Bayesian perspective, \cite{maleki2019flexible} considered a linear mixed effect model assuming that the random terms follow an unrestricted SN generalized-hyperbolic distribution, which provide flexibility for modeling complex data.

A common feature of these classes of LMMs is that the error terms are conditionally independent. However, in longitudinal studies, repeated measures are collected over time and hence the error term tends to be serially correlated.
There are some recent proposes in the literature that account for the time dependence in longitudinal data.
For instance, \cite{chang2016skew} proposed to use skew-normal antedependence models for modeling skewed longitudinal data exhibiting serial correlation, \cite{asar2018linear} proposed a methodology using multivariate normal variance-mean mixtures to fit linear mixed effects models for non-Gaussian continuous repeated measurement data, and \cite{lachos2019flexible} considered a robust generalization of the multivariate censored linear mixed model based on the scale mixtures of normal (SMN) distributions, with a damped exponential correlation (DEC) structure to take into account the autocorrelation among measurements. 

Nevertheless, to the best of our knowledge, there are no studies in the SMSN-LMM with serially correlated error structures, such as damped exponential correlation \citep[DEC,][]{munoz1992parametric} or autoregressive correlation of order $p$  \citep[AR($p$),][]{box1976time}. 
Therefore, the aim of this paper is to develop a full likelihood approach to SMSN-LMM with serially correlated errors, considering some useful correlation structures.
Our proposal intends to develop additional tools not considered in
\cite{Lachos_Ghosh_Arellano_2009} and to apply these techniques for making robust inferences in practical longitudinal data analysis. Moreover,  the proposed method has been coded and implemented in the \textsf{R} package \emph{skewlmm} \citep{skewlmm-manual}. A great advantage of this package is that it offers an automatic fit of all the SMSN-LMM taken into consideration.

The rest of the paper is organized as follows. Section \ref{sec:model} gives a brief introduction to SMSN class, further we define the SMSN-LMM and present some important dependence structures.
A likelihood approach for parameter estimation is given in Section \ref{sec:MLest}, including the estimation of random effects and standard errors.
Section \ref{sec:simstudies} presents some simulation studies that were conducted to evaluate the empirical performance of the proposed model under several scenarios and in Section \ref{sec:aplic} we fit the SMSN-LMM to a schizophrenia data set.
Finally, Section \ref{sec:conclude} presents some concluding remarks.

\section{Model formulation}\label{sec:model}

\subsection{Scale mixture of skew-normal distributions}\label{subsec:smsn}
%{\color{red} The SMSN class was motivated ...}
Let  $\Y$ be a $p\times 1$ random vector, $\bmu$ a $p\times 1$ location vector, $\bSigma$ a $p\times p$ positive definite dispersion matrix, $\blambda$ a $p \times 1$ skewness parameter, and let $U$ be a positive random variable with a cdf $H(u;\bnu)$, where $\bnu$ is a scalar or parameter vector indexing the distribution of $U$.
The multivariate SMSN class of distributions, denoted by $\textrm{SMSN}_p(\bmu,\bSigma,\mathbf{\blambda};H)$, can be defined through the following probability density function (pdf):
% \begin{equation}
% f(\mathbf{y})=2\int_{0}^{\infty}\phi_p(\mathbf{y}|\bmu,u^{-1}\bSigma)\
% \Phi(u^{1/2}\blambda^{\top}
% \bSigma^{-1/2}\left(\mathbf{y}-\bmu)\right)d
% H(u;\bnu),\,\,\,\,\,\mathbf{y}\in \mathbb{R}^p,\label{lsdefAB}
% \end{equation}
\begin{equation}
f(\mathbf{y})=2\int_{0}^{\infty}\phi_p(\mathbf{y};\bmu,\kappa(u) \bSigma)\
\Phi(\kappa(u)^{-1/2}\blambda^{\top}
\bSigma^{-1/2}\left(\mathbf{y}-\bmu)\right)d
H(u;\bnu),\,\,\,\,\,\mathbf{y}\in \mathbb{R}^p,\label{lsdefAB}
\end{equation}
for some positive weight function $\kappa(u)$, where $\phi_p(\cdot ;\bmu,\bSigma)$ denotes the pdf of the
$p$-variate normal distribution with a mean vector $\bmu$ and a
covariance matrix $\bSigma$, \, $\bSigma^{-1/2}$ is such that
$\bSigma^{-1/2}\bSigma^{-1/2}=\bSigma^{-1}$, and $\Phi(\cdot)$ denotes
the cumulative distribution function (cdf) of the standard normal
distribution.% When $\kappa(u)=u^{-1}$, the SMSN class reduces to the SNI class, as proposed in \cite{Lachos_Ghosh_Arellano_2009}. 

A special case of the SMSN class is
the skew--normal distribution \citep{azzalini96}, denoted by $\textrm{SN}_p(\bmu,\bSigma,\mathbf{\blambda})$, for which $H$ is
degenerate at $1$ (that is, $U=1$ with probability $1$), leading to the usual pdf
\begin{equation*}
    f(\mathbf{y})=2\phi_p(\mathbf{y};\bmu, \bSigma)
\Phi(A),\,\,\,\,\,\mathbf{y}\in \mathbb{R}^p,
\end{equation*}
where $A = \blambda^{\top}
\bSigma^{-1/2}\left(\mathbf{y}-\bmu\right)$. Another special case is obtained when the skewness parameter $\blambda=\mathbf{0}$, then the SMSN distribution in (\ref{lsdefAB}) reduces to the SMN distribution ($\Y\sim
\textrm{SMN}_p(\bmu,\bSigma;H)$), discussed earlier by
\cite{Lange93}. 

An important feature of this class, that can be used to derive many of its properties, is its stochastic representation. Let $\mathbf{Y}$ be a $p$--dimensional random vector with pdf as in \eqref{lsdefAB}, then $\mathbf{Y}$ can be represented in a stochastic way as follows:
\begin{equation}
\mathbf{Y}\buildrel d\over =\bmu+\kappa(U)^{1/2}\bSigma^{1/2}(\bdelta|T_0|+
(\mathbf{I}_p-\bdelta\bdelta^{\top})^{1/2}\mathbf{T}_1),\quad
\textrm{with} \quad
\bdelta=\frac{\blambda}{\sqrt{1+\blambda^{\top}\blambda}},\label{multi11}
\end{equation}
where $``\buildrel d\over ="$ means ``equal in distribution",
$|T_0|$ denotes the absolute value of $T_0$, $U\sim
H(\cdot;\bnu)$, $T_0\sim \textrm{N}_1(0,1)$, and $\mathbf{T}_1\sim
\textrm{N}_p(\mathbf{0},\mathbf{I}_p)$ are all independent
variables, with $\mathbf{I}_n$ being the $n\times n$ identity
matrix. The representation in (\ref{multi11}) facilitates the
implementation of EM-type algorithm. 
In this representation, it is straightforward that $\Y|U=u \sim \textrm{SN}_p(\bmu,\kappa(u)\bSigma,\mathbf{\blambda})$.

{Another useful feature of this class is that if $\mathbf{Y}\sim \textrm{SMSN}_p(\bmu,\bSigma,\mathbf{\blambda};H)$ and $\mathbf{X}\sim \textrm{SMN}_p(\bmu,\bSigma;H)$, then for any even function $g$, $g(\mathbf{Y}-\bmu)$ has the same distribution as $g(\mathbf{X}-\bmu)$ \citep{davila2018finite}. As a consequence, the Mahalanobis distance from the asymmetrical class  $d = \left(\mathbf{Y}-\bmu\right)^\top\bSigma^{-1}\left(\mathbf{Y}-\bmu\right)$  has the same distribution as the one from the symmetrical class $\left(\mathbf{X}-\bmu\right)^\top\bSigma^{-1}\left(\mathbf{X}-\bmu\right)$.
}

{For simplicity and following \cite{Lachos_Ghosh_Arellano_2009}, in the remaining of this work we restrict to the case where $\kappa(u)=u^{-1}$. }
The asymmetrical class of SMSN distributions includes many distributions as special cases, 
and we consider explicitly the following distributions:
%for example considering $\kappa(u)=u^{-1}$ we can derive
%such as the skew-normal (SN), the skew--$t$ (ST),the skew--slash (SL) and the skew--contaminated normal (SCN), as special cases. 
\begin{itemize}
\item[$\bullet$]{\it The multivariate skew--t distribution} with $\nu$ degrees of freedom, $\ST_p(\bmu,\bSigma,\blambda,\nu)$ \citep{Branco_Dey01,azzalini08}, which can be derived from the mixture model \eqref{lsdefAB} by taking $U\sim \textrm{Gamma}(\nu/2,\nu/2),$ with $\nu>0,$ and whose pdf can be written as
\begin{equation*}
    f(\mathbf{y})=2 t_p(\mathbf{y};\bmu, \bSigma,\nu)
T\left(\sqrt{\frac{\nu+p}{\nu+d}} A; \nu+p \right),\,\,\,\,\,\mathbf{y}\in \mathbb{R}^p,
\end{equation*}
where $t_p(\cdot;\bmu, \bSigma,\nu)$ and $T_p(\cdot;\nu)$ denote, respectively, the pdf of the $p$-variate Student-t distribution and the cdf of the standard univariate Student-t distribution with $\nu$ degrees of freedom, and $d = \left(\mathbf{y}-\bmu\right)^\top\bSigma^{-1}\left(\mathbf{y}-\bmu\right)$ is the Mahalanobis distance. {In this case, it can be shown that
$$d=\left(\mathbf{Y}-\bmu\right)^\top\bSigma^{-1}\left(\mathbf{Y}-\bmu\right) \sim p F(p,\nu),$$
where $F(a,b)$ denotes the Snedecor's F distribution with parameters $a$ and $b$.}

\item[$\bullet$]{\it The multivariate skew--slash distribution}, $\SSL_p(\bmu,\bSigma,\blambda,\nu)$, that arises by taking $U\sim \textrm{Beta}(\nu,1)$, with $u\in (0,1)$ and $\nu>0$, and whose pdf function takes the form
\begin{equation*}
    f(\mathbf{y})=2\nu\int_{0}^{1}u^{\nu-1} \phi_p(\mathbf{y};\bmu,u^{-1} \bSigma)\
\Phi(u^{1/2} A)du,\,\,\,\,\,\mathbf{y}\in \mathbb{R}^p.
\end{equation*}
{For this distribution, the cdf of the Mahalanobis distance is
$$P(d\leq r) = P\left(\chi_p^2\leq r \right) - \frac{2^\nu \Gamma(p/2+\nu)}{r^\nu \Gamma(p/2)} P\left(\chi_{p+2\nu}^2\leq r \right) .$$}
\item[$\bullet$]{\it The multivariate skew--contaminated normal distribution},
$\SCN_p(\bmu,\bSigma,\blambda,\nu_1,\nu_2),$ where $\nu_1,\nu_2\in (0,1)$ which arises when the mixing scale factor $\U$ is a discrete random variable taking one of two values and with
probability function given by  $h(u|\bnu)=\nu_1
\mathbb{I}_{\{\nu_2\}}(u)+(1-\nu_1) \mathbb{I}_{\{1\}}(u),$ where
$\bnu=(\nu_1,\nu_2)$ and $\mathbb{I}_{\{\tau\}}(u)$ is the indicator
function of the set $\tau$ whose value equals one if $u\in\tau$ and
zero elsewhere. In this case, the pdf becomes
\begin{equation*}
    f(\mathbf{y})=2\left\{\nu_1\phi_p(\mathbf{y};\bmu, \nu_2^{-1}\bSigma)
\Phi(\nu_2^{1/2}A) +(1-\nu_1)\phi_p(\mathbf{y};\bmu, \bSigma)
\Phi(A)  \right\},\,\,\,\,\,\mathbf{y}\in \mathbb{R}^p.
\end{equation*}
{It is easy to see that the cdf of the Mahalanobis distance in this case is given by
$$P(d\leq r) = \nu_1 P\left(\chi_p^2\leq \nu_2 r \right) + (1-\nu_1)P\left(\chi_p^2\leq r \right) .$$}

\end{itemize}

We refer to \cite{Lachos_Ghosh_Arellano_2009} and \cite{davila2018finite} for details and additional properties related to this class of distributions.

%%%%%%%%%%%%%%%%%%%%%%%%%%%%%%%%%%%%%%%%%%%%%%%%%%%%%%%%%%
\subsection{The SMSN-LMM}

Suppose that a variable of interest together with
several covariates are repeatedly measured for each of $n$
subjects at certain occasions over a period of time. 
For the $i$th subject, $i=1,\ldots,n$, let $\textbf{Y}_i$ be a $n_i\times 1$ vector of observed continuous responses. In general, a normal linear mixed effects model is defined as
\begin{equation}
\textbf{Y}_i=\textbf{X}_i\bbeta+\textbf{Z}_i\textbf{b}_i+\bepsilon_i,\,\,\,\,\ii\,,
\label{modeleq}
\end{equation}
where $\textbf{X}_i$ of dimension $n_i\times l$ is the design matrix corresponding to the fixed effects, $\bbeta$ of dimension $l\times 1$ is a vector of population-averaged regression coefficients called fixed effects, $\textbf{Z}_i$ of dimension $n_i\times q$  is the design matrix corresponding to the $q\times 1$ random effects vector $\textbf{b}_i$, and $\bepsilon_i$ of dimension $n_i\times 1$ is the vector of random errors. It is assumed that the random effects $\textbf{b}_i$ and the residual components $\bepsilon_i$ are
independent with $\textbf{b}_i\iid N_q(\mathbf{0},\textbf{D})$ and $\bepsilon_i\ind N_{n_i}(\mathbf{0},\bSigma_i)$. The $q \times q$  random effects covariance matrix $\textbf{D}$ may be unstructured or structured, and the $n_i \times n_i$ error covariance matrix $\bSigma_i$ is commonly written as $\sigma_e^2 \R_i$, where $\R_i$ can be a known matrix or a structured matrix depending on a vector of parameter, say $\bphi$.

%Considering hereafter $\kappa(u)=u^{-1}$, 
Likewise, the SMSN-LMM can be defined by considering
\begin{equation}\label{modSnmis}
    \left( \begin{array}{c}
         \textbf{b}_i \\
         \bepsilon_i
    \end{array}\right) \ind \textrm{SMSN}_{q+n_i}
    \left( \left(\begin{array}{c}
         c\bDelta \\
         \mathbf{0}
    \end{array} \right), \left(\begin{array}{cc}
         \mathbf{D} & \mathbf{0} \\
         \mathbf{0} & \bSigma_i
    \end{array} \right), \left(\begin{array}{c}
         \blambda \\
         \mathbf{0}
    \end{array} \right); H \right),\,\,\ii,
\end{equation}
where $c= c(\bnu)=-\sqrt{\frac{2}{\pi}}k_1$, with $k_1=\E\{U^{-1/2}\}$, $\bDelta = \textbf{D}^{1/2} \bdelta$, $\textbf{D} = \textbf{D}(\balpha)$ depends on unknown and reduced parameter vector $\balpha$, and we consider $\bSigma_i = \sigma_e^2 \R_i$, with $\R_i = \R_i(\bphi)$, $\bphi=(\phi_1,\hdots,\phi_p)^\top$, being a structured matrix. Calculating $k_1$ is straightforward and the results for the distributions discussed in Subsection \ref{subsec:smsn} are presented in Table \ref{tab:resdist}, where $h(\cdot;\bnu)$ is the pdf of $U$.

\begin{table}[ht]\centering
\caption{Summary of some important distributions from SMSN class.}
\label{tab:resdist}
\begin{tabular}{@{}cccc@{}}
\toprule
Distribution &$h(\cdot;\bnu)$ & $k_1 = \E\{U^{-1/2}\}$& $k_2 = \E\{U^{-1}\}$\\ \midrule
SN           &$1$& $1$& $1$\\
ST           & $\textrm{Gamma}(\nu/2,\nu/2)$& $\sqrt{\frac{\nu}{2}}\frac{\Gamma\left(\frac{\nu-1}{2}\right)}{\Gamma\left(\frac{\nu}{2}\right)}$,  \footnotesize if $\nu>1$ & $\frac{\nu}{\nu-2},$ \footnotesize if $\nu>2$ \\
SSL           &$\textrm{Beta}(\nu,1)$&$\frac{\nu}{\nu-1/2}$, \footnotesize if $\nu>1/2$& $\frac{\nu}{\nu-1}$, \footnotesize if $\nu>1$\\
SCN          &$\left\{\begin{array}{ll}
     \nu_2,& \textrm{\footnotesize w/ prob. } \nu_1 \\
    1, &  \textrm{\footnotesize w/ prob. } 1- \nu_1
\end{array} \right.$ & $1 + \nu_1\left(\nu_2^{-1/2}-1\right)$ &  $1 + \nu_1\left(\nu_2^{-1}-1\right)$ 
\\ 
\bottomrule
\end{tabular}
\end{table}

Some remarks about the model formulated in \eqref{modeleq} and \eqref{modSnmis} are worth noting: 
\begin{itemize}
    \item[i)] From \cite[Lemma 1 in Appendix A]{Lachos_Ghosh_Arellano_2009} it follows that, marginally, 
    \begin{equation}
         \textbf{b}_i\iid
 \textrm{SMSN}_q(c\bDelta,\textbf{D},\blambda;H)\,\,\,\textrm{and}\,\,\,
 \bepsilon_i\ind
 \textrm{SMN}_{n_i}(\mathbf{0},\sigma_e^2\textbf{R}_i;H),\,\,\ii.
    \end{equation}
    Thus the skewness parameter $\blambda$ incorporates asymmetry only in the distribution of the random effects (and consequently in the marginal distribution of $\Y$, which is given below). 
    In addition, as long as $k_1<\infty$ the chosen location parameter ensures that $\E\{\bi\} = \E\{\bepsilon_i\} = \mathbf{0}$, so that $\E\{\Y_i\} =\textbf{X}_i\bbeta$, for each $\ii$ and the regression parameter are all comparable.
    This is important since centering $\bi$'s distribution in $\mathbf{0}$ \citep{Lachos_Ghosh_Arellano_2009}, so that $\E\{\bi\} \neq \mathbf{0}$, might lead to biased estimates of $\bbeta$, as illustrated in Appendix \ref{apsec:extrasim}.%%%
    \item[ii)] Even though for each $\ii$, $\bi$ and $\bepsilon_i$ are indexed by the same scale mixing factor $U_i$ -- and hence they are not independent in general --, conditional on $U_i$, we have that $\bi$ and $\bepsilon_i$ are independent, what can be written as $\bi|U_i \perp \bepsilon_i|U_i$.
    Since $\textrm{Cov}\{\bi,\bepsilon_i\} = \E\{\bi\bepsilon_i^\top\} = \E_{U_i}\{\E\{\bi\bepsilon_i^\top|U_i\}\} = \mathbf{0}$, $\bi$ and $\bepsilon_i$ are uncorrelated.  Consequently, if $k_2 = \E\{U^{-1}\} < \infty$ we have 
    \begin{eqnarray}\label{eq:varY}
\nonumber\text{Var}(\Y_i) &=&\text{Var}(\bepsilon_i) + \textbf{Z}_i \text{Var}(\bi) \textbf{Z}_i^\top \\
    &=& k_2\left(\bSigma_i + \textbf{Z}_i \textbf{D} \textbf{Z}_i^\top \right) -  c^2\textbf{Z}_i \bDelta\bDelta^\top\textbf{Z}_i^\top = \bUpsilon_i. 
    \end{eqnarray}%%%
    \item[iii)] Under the SMSN-LMM at \eqref{modeleq} and \eqref{modSnmis},  for $\ii$, we have marginally
    \begin{equation}\label{eq:marginalY}
        \Y_i \ind \textrm{SMSN}_{n_i}(\textbf{X}_i\bbeta+\textbf{Z}_i c \bDelta,\bPsi_i,\bar{\blambda}_i;H),
    \end{equation}
    where $\bPsi_i = \bSigma_i + \textbf{Z}_i \textbf{D} \textbf{Z}_i^\top$,
    $\bar{\blambda}_i = \dfrac{\bPsi_i^{-1/2}\textbf{Z}_i\textbf{D}\bzeta}{\sqrt{1 + \bzeta^\top\bLambda_i\bzeta}}$, with $\bzeta=\textbf{D}^{-1/2}\blambda$ and $\bLambda_i=(\textbf{D}^{-1}+\textbf{Z}_i^\top \bSigma^{-1}_i\textbf{Z}_i)^{-1}$.
    Hence, the marginal pdf of $\Y_i$ is
    \begin{equation}\label{eq:marginalpdfY}
f(\mathbf{y}_i;\btheta)=2\int_{0}^{\infty}\phi_{n_i}(\mathbf{y}_i;\textbf{X}_i\bbeta+\textbf{Z}_i c \bDelta,u^{-1} \bPsi_i)
\Phi\left(u^{1/2}\bar{\blambda}_i^{\top}
\bPsi_i^{-1/2}\left(\mathbf{y}_i-\textbf{X}_i\bbeta-\textbf{Z}_i c \bDelta\right)\right)dH(u;\bnu).
\end{equation}
This result can be shown using arguments from conditional probability and the moment generating function of the multivariate skew-normal distribution, which is given in Appendix \ref{apsec:lemmas}.

    \item[iv)] The SMSN-LMM can be written hierarchically as follows:
    \begin{eqnarray}\label{eq:hierarq1}
    \Y_i|\bi,U_i=u_i &\ind& \textrm{N}_{n_i}\left(\textbf{X}_i\bbeta+\textbf{Z}_i\textbf{b}_i,
    u_i^{-1}\sigma_e^2\textbf{R}_i\right),\\
    \bi|T_i=t_i,U_i=u_i &\ind& \textrm{N}_{q}\left(\bDelta t_i,
    u_i^{-1}\bGamma\right),\\
    T_i|U_i=u_i &\ind& \textrm{TN}\left(c,u_i^{-1}, (c,\infty)\right), \,\textrm{ and }\\
    U_i &\ind&  \textrm{H}(\cdot;\bnu), \,\,\,\,\, \ii, \label{eq:hierarq4}
    \end{eqnarray}
    which are all independent, where $\bDelta = \textbf{D}^{1/2} \bdelta$, $\bGamma = \textbf{D}-\bDelta \bDelta^\top$, with $\bdelta = \blambda/\sqrt{1 + \blambda^\top \blambda}$ and $\textbf{D}^{1/2}$ is the square root of $\textbf{D}$, such that $\textbf{D}^{1/2}\textbf{D}^{1/2}=\textbf{D}$, containing $q(q+1)/2$ distinct elements, and TN$(\mu,\tau,(a,b))$ denotes the univariate normal distribution (N$(\mu,\tau)$) truncated on the interval $(a,b)$. The hierarchical representation given in (\ref{eq:hierarq1})-(\ref{eq:hierarq4}) is useful for the implementation of the EM algorithm as will be seen in the next section.
\end{itemize}

%%%%%%%%%%%%%%%%%%%%%%%%%%%%%%%%%%%%%%
\subsection{Within-subject dependence structures}\label{subsec:covstructs}
{%\color{red}
In order to enable some flexibility when modeling the error covariance, we consider essentially three dependence structures: conditionally independent, AR($p$) and DEC, which will be discussed next.\\

\noindent\emph{\it Conditional independence}\\
The most common and simplest approach is to assume that the error terms are conditionally independent (CI). Under this assumption, for each $\ii$, we have $\textbf{R}_i = \mathbf{I}_{n_i}$. This situation has been considered by \cite{Lachos_Ghosh_Arellano_2009} in their applications and will be denoted CI-SMSN-LMM.

In longitudinal studies, however, repeated measures are collected over time and hence the error term tends to be serially correlated.
In order to account for the within-subject serial correlation, we consider other two general structures.\\

\noindent\emph{\it Autoregressive dependence of order $p$}\\
Consider at first the case where for each subject $\ii$ a variable of interest is observed regularly over discrete time, $n_i$ times.
Then, we propose to model $\mathbf{R}_{i}$ as a structured AR($p$) dependence matrix \citep{box1976time}. Specifically,
\begin{equation}\label{riarp}
\mathbf{R}_{i}=\mathbf{R}_{i}(\bphi)=\frac{1}{1-\phi_1\rho_1-\ldots -\phi_p\rho_p}[\rho_{|r-s|}],
\end{equation}
where $r,s = 1, \ldots , n_i$ and $\rho_1,\hdots,\rho_p$ are the theoretical autocorrelations of the process, and thereby they are functions of autoregressive parameters $\bphi = (\phi_1,\ldots,\phi_p)^\top$, and satisfy the
Yule--Walker equations \citep{box1976time}, i.e.,
$$\rho_k=\phi_1\rho_{k-1}+\ldots+ \phi_p\rho_{k-p},\,\,\rho_0=1,\,\,\,k=1,\ldots,p.$$
In addition, the roots of $1-\phi_1B-\phi_2B^2-\cdots- \phi_pB^p=0$ must lie outside the unit circle to ensure stationarity of the AR($p$) model. %For the pure AR model, admissible values of $\bphi$ are confined in a $p$-dimensional hypercube $\mathbb{C}_p$.
Following \cite{barndorff1973parametrization}, the autoregressive process can be reparameterized using a one-to-one, continuous and differentiable transformation in order to simplify the conditions for stationarity.
For details on the estimation of the autoregressive coefficients, we refer to \cite{schumacher2017censored}.

The model formulated in \eqref{modeleq} and \eqref{modSnmis} with error covariance $\bSigma_i = \sigma_e^2 \textbf{R}_i$, where $\textbf{R}_i$ is given by \eqref{riarp}, $\ii$, will be denoted AR($p$)-SMSN-LMM.
{To accommodate situations in which measurements are taken irregularly over discrete time, we modify $\mathbf{R}_{i}$ by computing it for a regular range of time and then suppressing the line and column regarding the position from the missing measurements.}
\\

\noindent\emph{\it Damped exponential correlation}\\
More generally, consider now that for each subject $\ii$ a variable of interest is observed at times $\textbf{t}_i = (t_{i1},t_{i2},\hdots,t_{in_i})$. Following \cite{munoz1992parametric}, we propose to structure $\textbf{R}_i$ as a damped exponential correlation (DEC) matrix, as follows:
\begin{equation}\label{ridec}
    \textbf{R}_i = \textbf{R}_i(\bphi,\textbf{t}_i) = \left[ \phi_1^{|t_{ij}-t_{ik}|^{\phi_2}}\right], \,\,\,\, 0\leq\phi_1<1,\,\, \phi_2\geq 0,
\end{equation}
where $j,k=1,\hdots,n_i$, for $\ii$, and $\bphi=(\phi_1,\phi_2)^\top$.
Note that for $\phi_2=1$, $\textrm{R}_i$ reduces to the correlation matrix of a continuous-time autoregressive processes of order 1 (CAR($1$)), hence $\phi_2$ enables attenuation or acceleration  of the exponential decay from a CAR($1$) autocorrelation function, depending on its value. Moreover, $\phi_1$ describes the autocorrelation between observations such that $|t_{ij}-t_{ik}|=1$.
More details on DEC structure can be found in \cite{munoz1992parametric}.

The DEC structure is rather flexible, and some particular cases are worth pointing out:
\begin{enumerate}
    \item if $\phi_2=0$, then $\textbf{R}_i$ reduces to the compound symmetry correlation structure (CS);
    \item if $\phi_2=1$, then $\textbf{R}_i$ reduces to the CAR($1$) correlation structure;
    \item if $0<\phi_2<1$, then $\textbf{R}_i$ generates a decay rate slower than the CAR($1$) structure;
    \item if $\phi_2>1$, then $\textbf{R}_i$ generates a decay rate faster than the CAR($1$) structure; and
    \item if $\phi_2 \rightarrow \infty$, then $\textbf{R}_i$ converges to the correlation matrix of a moving-average of order 1 (MA(1)).
\end{enumerate}

The model formulated in \eqref{modeleq} and \eqref{modSnmis} with error covariance $\bSigma_i = \sigma_e^2 \textbf{R}_i$, where $\textbf{R}_i$ is given by \eqref{ridec}, $\ii$, will be denoted DEC-SMSN-LMM.
}

\subsection{The likelihood function}
The marginal pdf of $\Y_i$, $\ii$, is given in \eqref{eq:marginalpdfY}, with 
$\textbf{R}_i$ depending on the chosen correlation structure, as described in Subsection \ref{subsec:covstructs}.
Hence, the log-likelihood
function for $\btheta$ based on the observed sample
$\mathbf{y} = (\mathbf{y}^{\top}_1,\ldots,\mathbf{y}^{\top}_n)^{\top}$
is given by $$\ell(\btheta|\mathbf{y})=\sumas
\ell_i(\btheta|\mathbf{y})=\sumas
\log{(f(\mathbf{y}_i|\btheta))},$${ where }
$\btheta=(\bbeta^{\top}, \sigma_e^2, \bphi^{\top}, \balpha^{\top}, \blambda^{\top}, \bnu^\top)^{\top}$.
As the observed log-likelihood function
involves complex expressions, it is very difficult to work directly with $\ell(\btheta|\mathbf{y})$ to find the ML estimates of $\btheta$.
Thus, in this work we propose to use an EM-type algorithm \citep{Dempster77} for parameter estimation via ML.

\section{Maximum likelihood estimation}\label{sec:MLest}
\subsection{The EM algorithm}
%ecm para normal, ecme para outras
A convenient feature of the SMSN-LMM is its hierarchical representation, as given in \eqref{eq:hierarq1}--\eqref{eq:hierarq4}.
Following \cite{Lachos_Ghosh_Arellano_2009}, $\mathbf{b}$, $\mathbf{u}$ and $\mathbf{t}$ can be treated as hypothetical missing data and therefore we propose to use the ECME algorithm \citep{Liu94} for parameter estimation.

Let the augmented data set be 
$\mathbf{y}_c = (\yp^{\top}, \be^{\top}, \mathbf{u}^{\top},
\mathbf{t}^{\top})^{\top}$, where
$\yp = (\yp^{\top}_1, \ldots, \yp^{\top}_n)^{\top}$,
$\mathbf{b} = (\mathbf{b}^{\top}_1, \ldots, \mathbf{b}^{\top}_n)^{\top}$, $\mathbf{u} = (u_1, \ldots, u_n)^{\top}$
and $\mathbf{t} = (t_1, \ldots, t_n)^{\top}$. Hence, an
EM-type algorithm can be applied to the complete-data log-likelihood
function $\ell_c(\btheta|\mathbf{y}_c)=\sumas
\ell_i(\btheta|\mathbf{y}_c) $, given by
\begin{eqnarray*}\label{logcompleta}
	\ell_c(\btheta|\mathbf{y}_c)&=&\sumas\left[-\frac{1}{2}\log{|\sigma_e^2\textbf{R}_i|}
	-\frac{u_i}{2\sigma_e^2}(\yp_i-\mathbf{X}_i\bbeta
	-\mathbf{Z}_i\mathbf{b}_i)^{\top}\textbf{R}_i^{-1}
	(\yp_i-\mathbf{X}_i\bbeta-\mathbf{Z}_i\mathbf{b}_i)\right.\nonumber\\
	&&\left.-\frac{1}{2}\log{|\bGamma|}-
	\frac{u_i}{2}(\mathbf{b}_i-\bDelta
	t_i)^{\top}\bGamma^{-1}(\mathbf{b}_i- \bDelta t_i)\right]+K(\bnu)+C,
\end{eqnarray*}
where $C$ is a constant that is independent of the parameter vector $\btheta$ and $K(\bnu)$ is a function that depends on $\btheta$ only through $\bnu$.

%%%%
For the current value $\btheta=\widehat{\btheta}^{(k)}$, the E-step of
the EM-type algorithm requires the evaluation of
$\qhatk(\btheta) = \E\left\{\ell_c(\btheta|\mathbf{y}_c)\mid \y, \widehat{\btheta}^{(k)}\right\}=\sum^n_{i=1} \qhatk_i(\btheta)$,
where the expectation is taken with respect to the joint
conditional distribution of $\mathbf{b}$, $\mathbf{u}$ and
$\mathbf{t}$, given $\mathbf{y}$ and $\widehat{\btheta}$. Thus, we
have
$$\qhatk_i(\btheta)=\qhatk_{1i}(\bbeta,\sigma_e^2,\bphi)+
\qhatk_{2i}(\balpha,\blambda) + \qhatk_{3i}(\bnu),$$ {where}
% \begin{eqnarray*}
% \qhatk_{1i}(\bbeta,\sigma_e^2,\bphi) &=& -\frac{n_i}{2}\log\left(\widehat{\sigma}_e^{2(k)}\right) -\frac{1}{2}\log \left|\widehat{\textbf{R}}^{(k)}_i\right|
% -\frac{\widehat{u}_i^{(k)}}{2\widehat{\sigma}_e^{2(k)}}\left(\y_{i}-\X_i\widehat{\bbeta}^{(k)}\right)^{\top}\widehat{\textbf{R}}_i^{-1(k)}%^{-1}
% \left(\y_{i}-\X_i\widehat{\bbeta}^{(k)}\right)\nonumber\\&&+\frac{1}{\widehat{\sigma}_e^{2(k)}}\left(\y_{i}-\X_i\widehat{\bbeta}^{(k)}\right)^{\top}\widehat{\textbf{R}}_i^{-1(k)}\z_i\widehat{\mathbf{ub}}_i^{(k)}
% -\frac{1}{2\widehat{\sigma}_e^{2(k)}}\textrm{tr}\left(\widehat{\textbf{R}}_i^{-1(k)}\z_i\widehat{\mathbf{ub^2}}_i^{(k)}\z^{\top}_i\right),\,\,\,\,
% \\%\label{eq:Q1mat}\\
% \qhatk_{2i}(\balpha,\blambda) &=& -\frac{1}{2}\log{\left| \widehat{\bGamma}^{(k)} \right|}-\frac{1}{2}\textrm{tr}\left(\widehat{\bGamma}^{-1(k)}\widehat{\mathbf{ub^2}}_i^{(k)}\right)+\widehat{\bDelta}^{\top (k)}\widehat{\bGamma}^{-1(k)}\widehat{\mathbf{utb}}_i^{(k)}\nonumber
% \\ &&-\frac{\widehat{ut^2}_i^{(k)}}{2}\widehat{\bDelta}^{\top (k)}\widehat{\bGamma}^{-1(k)}\widehat{\bDelta}^{(k)}%\label{eq:Q2mat}
% ,
% \end{eqnarray*}
\begin{eqnarray*}
\qhatk_{1i}(\bbeta,\sigma_e^2,\bphi) &=& -\frac{n_i}{2}\log\left({\sigma}_e^{2}\right) -\frac{1}{2}\log \left|{\textbf{R}}_i\right|
-\frac{\widehat{u}_i^{(k)}}{2{\sigma}_e^{2}}\left(\y_{i}-\X_i{\bbeta}\right)^{\top}{\textbf{R}}_i^{-1}%^{-1}
\left(\y_{i}-\X_i{\bbeta}\right)\nonumber\\&&+\frac{1}{{\sigma}_e^{2}}\left(\y_{i}-\X_i{\bbeta}\right)^{\top}{\textbf{R}}_i^{-1}\z_i\widehat{\mathbf{ub}}_i^{(k)}
-\frac{1}{2{\sigma}_e^{2}}\textrm{tr}\left({\textbf{R}}_i^{-1}\z_i\widehat{\mathbf{ub^2}}_i^{(k)}\z^{\top}_i\right),\,\,\,\,
\\%\label{eq:Q1mat}\\
\qhatk_{2i}(\balpha,\blambda) &=& -\frac{1}{2}\log{\left| {\bGamma} \right|}-\frac{1}{2}\textrm{tr}\left({\bGamma}^{-1}\widehat{\mathbf{ub^2}}_i^{(k)}\right)+{\bDelta}^{\top }{\bGamma}^{-1}\widehat{\mathbf{utb}}_i^{(k)} %\nonumber \\ &&
-\frac{\widehat{ut^2}_i^{(k)}}{2}{\bDelta}^{\top}{\bGamma}^{-1}{\bDelta}%\label{eq:Q2mat}
,
\end{eqnarray*}
with $\textrm{tr}(\A)$ and $|\A|$ indicating trace and determinant of matrix $\A$, respectively, %$\widehat{\textbf{R}}^{(k)}_i = \textbf{R}_i(\widehat{\bphi}^{(k)})$, 
$\widehat{u}_i^{(k)} = \E\{U_i\mid \widehat{\btheta}^{(k)},\y_i\}$,
$\widehat{\mathbf{u}\be}_i^{(k)} = \E\{U_i \be_i\mid \widehat{\btheta}^{(k)},\y_i\}$,
$\widehat{\mathbf{ub^2}}_i^{(k)} = \E\{U_i \be_i\be_i^\top\mid \widehat{\btheta}^{(k)},\y_i\}$,
$\widehat{\mathbf{utb}}_i^{(k)} = \E\{U_i T_i \be_i\mid \widehat{\btheta}^{(k)},\y_i\}$,
$\widehat{ut}_i^{(k)} = \E\{U_i T_i\mid \widehat{\btheta}^{(k)},\y_i\}$,
and $\widehat{ut^2}_i^{(k)} = \E\{U_i T_i^2\mid \widehat{\btheta}^{(k)},\y_i\}$,
$\ii$.
These expressions can be readily evaluated once we have the following conditional distributions, which can be derived using arguments from conditional probability:
\begin{eqnarray}
\be_i |t_i,u_i,\y_i,\btheta &\sim& \textrm{N}_q(\s_i t_i+\br_i, u_i^{-1}\B_i),\nonumber\\
T_i|u_i,\y_i,\btheta &\sim& \textrm{TN}\left(c+\mu_i, u_i^{-1}M_i^2; (c,\infty)\right),\\
\Y_i|\btheta &\sim& \textrm{SMSN}_{n_i}(\X_i\bbeta+c \Z_i \bDelta, \bPsi_i, \bar{\blambda}_i;H), \nonumber
\end{eqnarray}
where $M_i=(1+\bDelta^\top \Z_i^\top \bOmega_i^{-1} \Z_i \bDelta)^{-1/2}$, $\mu_i=M_i^2\bDelta^\top \Z_i^\top \bOmega_i^{-1}(\y_i-\X_i\bbeta-c \Z_i \bDelta)$, $\B_i = (\bGamma^{-1}+\Z_i^\top \bSigma^{-1}_i\Z_i)^{-1}$, $\s_i=(\mathbf{I}_q-\B_i \Z_i^\top \bSigma^{-1}_i\Z_i)\bDelta$, $\br_i=\B_i \Z_i^\top \bSigma^{-1}_i (\y_i-\X_i\bbeta)$ , $\bOmega_i = \bSigma_i + \Z_i \bGamma \Z_i^\top$, for $\ii$.

Thence, after some algebra and omitting the supra-index $(k)$, we get the following expressions:
\begin{equation}%\small
    \begin{array}{l}
         \widehat{ut}_i = (\widehat{\mu}_i+\widehat{c})\ui + \widehat{M}_i \taui, \,\,
          \widehat{ut^2}_i = \widehat{M}_i^{2} + (\widehat{\mu}_i+\widehat{c})^2\ui + \widehat{M}_i(\widehat{\mu}_i+2\widehat{c}) \taui,\\
          \widehat{\mathbf{u}\be}_i = \widehat{\br}_i\ui + \widehat{\s_i} \widehat{ut}_i, \,\,
          \widehat{\mathbf{utb}}_i = \widehat{\br}_i\widehat{ut}_i + \widehat{\s_i} \widehat{ut^2}_i,\\
          \widehat{\mathbf{ub^2}}_i = \widehat{\B}_i + \ui \widehat{\br}_i\widehat{\br}_i^\top+ \widehat{ut}_i(\widehat{\s}_i\widehat{\br}_i^\top + \widehat{\br}_i\widehat{\s}_i^\top) + \widehat{ut^2}_i \widehat{\s}_i\widehat{\s}_i^\top,
    \end{array}
\end{equation}
where $\widehat{c} = c(\widehat{\bnu})$, and the expressions for $\ui$ and $\taui = \E\left\{ U_i^{1/2} \textrm{W}_\Phi\left(U_i^{1/2}{A}_i\right)\mid \widehat{\btheta},\y_i \right\} $, for $\textrm{W}_\Phi(x) = \phi_1(x)/\Phi(x)$, $x \in \mathbb{R}$ and $A_i = \mu_i/M_i = \bar{\blambda}_i^\top \bPsi_i^{-1/2}(\y_i - \X_i\bbeta-c \Z_i \bDelta)$, 
can be found in Section 2 from \cite{Lachos_Ghosh_Arellano_2009}, which can be easily implemented for the ST and SCN distributions, but involve numerical integration for the SSL case. 

%%% conditional maximization
The M-step requires the maximization of
$\qhatk(\btheta)$ with respect to $\btheta$. The
motivation for employing an EM-type algorithm is that it can be
utilized efficiently to obtain closed-form equations for the M-step.
The conditional maximization (CM) step conditionally maximize
$\qhatk(\btheta)$ with respect to $\btheta$,
obtaining a new estimate  $\widehat{\btheta}^{(k+1)}$, as follows:

%$\widehat{u}_i^{(k)}$,$\widehat{\mathbf{u}\be}_i^{(k)}$,
%$\widehat{\mathbf{ub^2}}_i^{(k)}$,
%$\widehat{\mathbf{utb}}_i^{(k)}$,
%$\widehat{ut}_i^{(k)}$,and $\widehat{ut^2}_i^{(k)}$,
\noindent{\bf 1.} Update
$\widehat{\bbeta}^{(k)}$, $\widehat{\sigma^2}^{(k)}_e$, $\widehat{\bphi}^{(k)}$, $\widehat{\bDelta}^{(k)}$ and $\widehat{\bGamma}^{(k)}$ using the following  expressions:
\begin{eqnarray*}
\widehat{\bbeta}^{(k+1)}&=&\left(\sumas\uki
\X_i^{\top}\widehat{\bSigma}^{{-1(k)}}_i\X_i\right)^{-1}\sumas
\X_i^{\top}\widehat{\bSigma}^{{-1(k)}}_i\left(\uki\yp_i-\Z_i
\widehat{\mathbf{u}\be}_i^{(k)}\right),\\
\widehat{\sigma^2_e}^{{(k+1)}}&=&\frac{1}{N}\sumas
\left[\uki\left(\yp_i-\mathbf{X}_i\widehat{\bbeta}^{(k+1)}\right)^{\top}\mathbf{R}_i^{-1}\left(\widehat{\bphi}^{(k)}\right)
\left(\yp_i-\mathbf{X}_i\widehat{\bbeta}^{(k+1)}\right)\right.\\
&&\left. -2 \left(\yp_i-\mathbf{X}_i\widehat{\bbeta}^{(k+1)}\right)^{\top}
\mathbf{R}_i^{-1}\left(\widehat{\bphi}^{(k)}\right)\Z_i\widehat{\mathbf{u}\be}_i^{(k)} 
+{\textrm{tr}}\left(\R_i^{-1}\left(\widehat{\bphi}^{(k)}\right)\z_i\widehat{\mathbf{ub^2}}_i^{(k)} \z^{\top}_i\right)\right],\\
\widehat{\bphi}^{(k+1)}&=&\underaccent{\bphi}{\textrm{argmax}}\sumas\left(-\frac{1}{2}\log|\textbf{R}_i(\bphi)|
-\frac{\uki}{2\widehat{\sigma}_e^{2(k+1)}}\left(\y_{i}-\X_i\widehat{\bbeta}^{(k+1)}\right)^{\top}\textbf{R}_i^{{-1}}(\bphi)
\left(\y_{i}-\X_i\widehat{\bbeta}^{(k+1)}\right)\right.\\&&\left.+
\frac{1}{\widehat{\sigma}_e^{2(k+1)}}\left(\y_{i}-\X_i\widehat{\bbeta}^{(k+1)}\right)^{\top} \textbf{R}_i^{-1}(\bphi) \z_i \widehat{\mathbf{u}\be}_i^{(k)}
-\frac{1}{2\widehat{\sigma}_e^{2(k+1)}}\textrm{tr}\left(\textbf{R}_i^{-1}(\bphi)\z_i\widehat{\mathbf{ub^2}}_i^{(k)}\z^{\top}_i\right)\right),\\
\widehat{\bDelta}^{(k+1)}&=&%\displaystyle
\frac{\sumas
	\widehat{\mathbf{utb}}_i^{(k)}}{\sumas \widehat{ut^2}_i^{(k)}},\\
\widehat{\bGamma}^{(k+1)}&=&\frac{1}{n}\sumas
	\left(\widehat{\mathbf{ub^2}}_i^{(k)}-
	\widehat{\mathbf{utb}}_i^{(k)} \widehat{\bDelta}^{{(k+1)\top}}
	-\widehat{\bDelta}^{(k+1)}	\widehat{\mathbf{utb}}_i^{(k)\top}
	+\widehat{ut^2}_i^{(k)}\widehat{\bDelta}^{(k+1)}\widehat{\bDelta}^{{(k+1)\top}}\right),
\end{eqnarray*}
where $N = \sumas n_i$. %escrever como é atualizacao de phi

\noindent{\bf 2.} Update $\widehat{\bnu}^{(k)}$ by
optimizing the constrained actual marginal log-likelihood
function $$\label{Q33esti}
\widehat{\bnu}^{(k+1)}=\underaccent{\bnu}{\textrm{argmax}} \{f(\yp;\widehat{\btheta}^{*(k+1)},\bnu)\},$$
where $f(\yp;\btheta)$ is as in \eqref{eq:marginalpdfY} and $\btheta^{*} = \btheta \setminus \bnu$.  

The skewness parameter vector and the
parameters from the scale matrix of random effects $\mathbf{b}$,
can be estimated by 
$$\widehat{\mathbf{D}}^{(k+1)}=\widehat{\bGamma}^{(k+1)}+\widehat{\bDelta}^{(k+1)}\widehat{\bDelta}^{(k+1)\top}$$
and
$$\widehat{\blambda}^{(k+1)}=\displaystyle \widehat{\mathbf{D}}^{-1/2 (k+1)}\widehat{\bDelta}^{(k+1)}/
(1-\widehat{\bDelta}^{{(k+1)\top}}\widehat{\mathbf{D}}^{-1(k+1)}\widehat{\bDelta}^{(k+1)})^{1/2}.$$ %obs about transformation in AR case
The algorithm is iterated until a predefined criteria is reached, such as when $\small \left|\ell(\widehat{\btheta}^{(k+1)}\mid \y)/\ell(\widehat{\btheta}^{(k)}\mid \y)-1\right|$ becomes small enough.

In practice, to select between various SMSN-LMM distributions we can consider the Akaike information criterion (AIC) and the Bayesian information criterion (BIC) \citep{wit2012all}, given by
\begin{eqnarray*}
\textrm{AIC} &=& -2\ell(\widehat{\btheta}) + 2 m, \,\,\textrm{ and } \\
\textrm{BIC} &=& -2\ell(\widehat{\btheta}) + m \log(N) ,
\end{eqnarray*}
where $m$ is the number of estimated parameters and $\widehat{\btheta}$ is the ML estimate of $\btheta$.

\subsection{Estimation of random effects and prediction}\label{subsec:prediction}

The minimum mean-squared error (MSE) estimator of $\mathbf{b}_i$
	is obtained by the conditional mean of $\mathbf{b}_i$ given $\Y_i=\y_i$, that can be shown to be
	\begin{eqnarray}\label{eq:estbi}
	\widehat{\be}_i(\btheta)&=&\E\{\mathbf{b}_i|\Y_i=\y_i,\btheta\}=\E_{U_i}\{\E\{\mathbf{b}_i|U_i,\Y_i=\y_i,\btheta\}|\Y_i=\y_i,\btheta\}\nonumber\\
	&=&\bmu_{bi}+\frac{\tau_{-1i}}{\sqrt{1+\bzeta^{\top}\bLambda_i\bzeta}}\,\bLambda_i\bzeta,
	\end{eqnarray}
 where $\bmu_{bi}=c\bDelta+\mathbf{D}\Z^{\top}_i\bPsi^{-1/2}_i\y_{0i}$, with
	$\y_{0i}=\bPsi^{-1/2}_i(\y_i-\X_i\bbeta-c\Z_i\bDelta)$, and $\bLambda_i$,
	$\bzeta$ and $\bar{\blambda}_{i}$ are as in \eqref{eq:marginalpdfY}. Explicit expression for 
	$\tau_{-1i}=\E\{U_i^{-1/2}W_{\Phi}(U_i^{1/2} A_i)\mid\y_i\}$, where 
	$A_i=\bar{\blambda}^{\top}_{i}\y_{0i}$, can be found in \cite{Lachos_Ghosh_Arellano_2009}, as well as the proof of the result, which comes from the fact that the conditional distribution of $\be_i$ given $(\Y_i,U_i)=(\y_i,u_i)$ belongs to the extended skew-normal (EST) family of distributions \citep{azzalini1999statistical}. 
In practice, the estimator of $\mathbf{b}_i$ -- also known as empirical Bayes estimator --,
$\widehat{\mathbf{b}}_i$, can be obtained by substituting the ML
estimate $\widehat{\btheta}$ into \eqref{eq:estbi}.

Furthermore, in practical applications it is usual the interest in predicting $\Y^+_i$, a future
$n_{\textrm{pred}}\times 1$ vector of measurement of $\Y_i$, given
the observed measurement
$\y=(\y^{\top}_{(i)},\y^{\top}_i)^{\top}$,
where
$\y_{(i)}=(\y^{\top}_1,\ldots,\y^{\top}_{i-1},\y^{\top}_{i+1},\ldots,\y^{\top}_{n})$.
If $\X^+_i$ and $\Z^+_i$ denote $n_{\textrm{pred}}\times p$ and
$n_{\textrm{pred}}\times q$ matrices of prediction regression variables corresponding
to $\Y^+_i$, we assume that
$$\left[\begin{array}{c}
\Y_i \\
\Y_i^+
\end{array}\right]\sim
\SMSN_{n_i+n_\textrm{pred}}(\X^*_i\bbeta,\bPsi^*_i,\bar{\blambda}_{i}^*;H),
$$
where $\X^*_i=(\X^{\top}_i,\X^{+\top}_i)^{\top}$,
$\Z^*_i=(\Z^{\top}_i,\Z^{+\top}_i)^{\top}$,
$\bPsi^*_i=\bSigma^*_i+\mathbf{Z}^*_i\mathbf{D}\mathbf{Z}_i^{*\top},$
$\bLambda^*_i=(\mathbf{D}^{-1}+\mathbf{Z}_i^{*\top}\bSigma^{*-1}_i
\mathbf{Z}_i^*)^{-1},$ $\bar{\blambda}^{*}_{i}=
\dfrac{\bPsi_i^{*-1/2}\mathbf{Z}_i^*\mathbf{D}\bzeta}
{\sqrt{1+\bzeta^{\top}\bLambda^*_i\bzeta}}$, and $\bPsi^*_i =\small \left(\begin{array}{cc}
     \bPsi_{i 11}^*&\bPsi_{i 12}^*  \\
     \bPsi_{i 21}^*&\bPsi_{i 22}^* 
\end{array}\right)$. 
From \cite{Lachos_Ghosh_Arellano_2009}, it follows that the minimum MSE predictor of future measurements of $\Y_i$ is the conditional expectation of $\yp_i^+$ given $\Y_i = \y_i$, i.e.,
\begin{eqnarray}
\widehat{\Y}^+_i(\btheta)&=&\E\{\Y^+_i|\mathbf{Y}_i= \y_i,\btheta\}
=\E_{U_i}\{\E\{\Y^+_i|U_i,\mathbf{Y}_i= \y_i\}|\mathbf{Y}_i= \y_i,\btheta\} \nonumber\\
&=&\bmu_{i2.1}+\frac{\widetilde{\tau}_{-1i}\bPsi^*_{i22.1}\bupsilon^{(2)}_i}{\sqrt{1+\bupsilon^{(2)\top}_i\bPsi^*_{i22.1}\bupsilon^{(2)}_i}},\label{eq:ycond}
\end{eqnarray}
where
$\bmu_{i2.1}=\X^+_i\bbeta+c\Z_i^+\bDelta+\bPsi^*_{i21}\bPsi^{*-1}_{i11}(\y_i-\X_i\bbeta-c \Z_i\bDelta),
\bPsi^*_{i22.1}=\bPsi^*_{i22}-\bPsi^*_{i21}\bPsi^{*-1}_{i11}\bPsi^*_{i12},$
$\bupsilon_i=\bPsi^{*-1/2}_i\bar{\blambda}_{i}^*=(\bupsilon^{(1)\top}_i,\bupsilon^{(2)\top}_i)^{\top}$, $\bPsi^{*}_{i11}=\bPsi_i$,
$\bPsi^*_{i12}=\bPsi^{*\top}_{i21}$, and  
$$\widetilde{\tau}_{-1i}=\E\left\{U_i^{-1/2}W_{\Phi}\left(U_i^{1/2} \widetilde{\bupsilon}_i^\top (\Y_i-\X_i\bbeta-c\Z_i\bDelta)\right)\mid\Y_i=\y_i\right\},$$ with $\widetilde{\bupsilon}_i =\left({\bupsilon^{(1)}_i + \bPsi^{*-1}_{i11}\bPsi^*_{i12}\bupsilon^{(2)}_i}\right)/{\sqrt{1+\bupsilon^{(2)\top}_i\bPsi^*_{i22.1}\bupsilon^{(2)}_i}}$.

In practice, the prediction of $\Y^+_i$ can be obtained by substituting the ML
estimate $\widehat{\btheta}$ into \eqref{eq:ycond}, so
$\widehat{\Y}^+_i=\widehat{\Y}^+_i(\widehat{\btheta})$.

\subsection{Estimation of standard errors (SE)}\label{sec:stderror}
In this section, we derive the observed information matrix from the score vector with respect to $\btheta^* = \btheta\setminus\bnu$.
First, we reparameterize $\mathbf{D}=\mathbf{F}^2$ for ease of computation and
theoretical derivation, where $\mathbf{F}$ is the square root of
$\mathbf{D}$ containing $q(q + 1)/2$ distinct elements
$\balpha_b=(\alpha_1,\ldots,\alpha_{q(q + 1)/2})^{\top}$. Given the
observed sample
$\mathbf{y}=(\mathbf{y}^{\top}_1,\ldots,\mathbf{y}^{\top}_n)^{\top}$ and $\bnu$,
the log-likelihood function for
$\btheta^*=(\btheta^{\top}_1,\btheta^{\top}_2)^{\top}$,
with $\btheta_1=(\bbeta^{\top},\sigma^2_e,\bphi^{\top})^{\top}$ and
$\btheta_2=(\balpha^{\top}_{b},\blambda^\top)^{\top}$ is given by
$\ell(\btheta^*)=\sumas \ell_i(\btheta^*;\bnu)$, where
\begin{equation}\label{logvero}
\ell_i(\btheta^*;\bnu)=\log{2}-\displaystyle\frac{n_i}{2}log{2\pi}-\frac{1}{2}\log{|\bPsi_i|}+\log{K_i},
\end{equation}
with
$$K_i(\btheta^*;\bnu)=\int^{\infty}_{0}u_i^{n_i/2}\exp\left\{-\frac{1}{2}u_i\rm{d}_i\right\}\Phi\left(u^{1/2}_iA_i\right)dH(u_i;\bnu),$$
where
$$\rm{d}_i=(\mathbf{y}_i-\X_i\bbeta-c\Z_i\bDelta)^{\top}\bPsi_i^{-1}(\mathbf{y}_i-\X_i\bbeta-c\Z_i\bDelta) \,\, \textrm{ and } \,\, A_i=\frac{\blambda^{\top}\mathbf{F}\Z_i\bPsi^{-1}_i(\yp_i-\X_i\bbeta-c\Z_i\bDelta)}{(1+\blambda^{\top}\mathbf{F}^{-1}\bLambda_i\mathbf{F}^{-1}\blambda)^{1/2}}.$$
Thus, we have after some algebraic manipulations that the score
vector is given by
\begin{equation}\label{eq:score}
\mathbf{s}=\sumas \mathbf{s}_i =\sumas
\frac{\partial\ell_i(\btheta^*;\bnu)}{\partial\btheta^*}=-\frac{1}{2}\sumas\frac{\partial\log|\bPsi_i|}{\partial\btheta^*}+\sumas
\frac{1}{K_i}\frac{\partial K_i}{\partial\btheta^*},
\end{equation}
where \,
$\displaystyle\frac{\partial
	K_i}{\partial\btheta^*}=I^{\phi}_i\left(\frac{n_i+1}{2}\right)\displaystyle\frac{\partial
	A_i}{\partial\btheta^*}-\frac{1}{2}I^{\Phi}_i\left(\frac{n_i+2}{2}\right)\displaystyle\frac{\partial
	\rm d_i}{\partial\btheta^*}$, with
\begin{eqnarray}
I^{\Phi}_i(w)=\int^{\infty}_{0}u_i^{w}\exp\left\{-\displaystyle\frac{1}{2}u_i\rm{d}_i\right\}\Phi\left(u^{1/2}_iA_i\right)dH(u_i;\bnu),\label{eq:IPhi}\\
I^{\phi}_i(w)=\int^{\infty}_{0}u_i^{w}\exp\left\{-\displaystyle\frac{1}{2}u_i\rm{d}_i\right\}\phi_1\left({u^{1/2}_iA_i}\right)dH(u_i;\bnu),\label{eq:Iphi}
\end{eqnarray}
and $K_i=I^{\Phi}_i(\frac{n_i}{2})$. 
The results from substituting $H$ for each distribution considered are presented in Appendix \ref{apsec:infmat},
along with the derivatives of $\log{|\bPsi_i|}$, $\rm d_i$ and $A_i$, which involve tedious but not complicated algebraic manipulations.

Under some regularity conditions, asymptotic covariance of ML estimates can be estimated by the inverse of the observed
information matrix, $\textrm{I}(\widehat{\btheta^*}) = \sumas \widehat{\bf s}_i \widehat{\bf s}_i^\top$, where $ \widehat{\bf s}_i = {\bf s}_i(\widehat{\btheta^*})$ is
the score vector in \eqref{eq:score} evaluated at $\btheta^*=\widehat{\btheta^*}$.

\subsection{Likelihood ratio test}\label{subsec:lrt}

Considering the usual interest in testing if a restricted model represents the data well enough, in this section we present a likelihood ratio test. An important particular case is testing the hypothesis that an asymmetrical model is not necessary, that could be written as $\Ha_0: \blambda = \textbf{0}$.

Let $\Ha_0: \btau=\mathbf{0}$, $\btau=(\tau_1,\hdots,\tau_r)^\top$, be a hypothesis of interest, $\bTheta$ be the $k$-dimensional parameter space of the unrestricted model, and $\bTheta_0$ be the parameter space under $\Ha_0$, for $1\leq r<k$.
With the interest of measuring the impact of $\Ha_0$ in the maximum of the likelihood function, 
consider the statistic
$\Lambda_n = 2\left(\ell(\widehat{\btheta}) - \ell(\widehat{\btheta}_0) \right)$, where $\widehat{\btheta}_0$ is the ML estimate of $\btheta$ under the restriction in $\Ha_0$.
Then, under $\Ha_0$, $\Lambda_n$ is asymptotic distributed as a chi-square random variable with $r$ degrees of freedom ($\chi^2_r$) \citep{mood1950introduction}.

\subsection{Additional tools for model evaluation}\label{subsec:residuals}
{Evaluating the suitability of a fitted model to a real data set is an important step in data analysis and there are several methods that can be used to this purpose. 
When dealing with heavy-tailed data, the Mahalanobis distance is a convenient measure which can be used to identify potential outlying observations and to assess the validity of the underlying distributional assumption of the response variable, once if the fitted model is appropriate the distribution of the Mahalanobis distance is known and presented in Subsection \ref{subsec:smsn}. 

Following \cite{ho2010robust}, to assess the goodness of fit of SMSN-LMM one can construct a Healy-type plot \citep{healy1968multivariate} by plotting the nominal probability values $1/n, 2/n, \hdots, n/n$ against the theoretical cumulative probabilities of the ordered observed Mahalanobis distances. If the fitted model is appropriate, the plot should resemble a straight line through the origin with unit slope. 

Additionally, based on \cite{zeller2010influence}, the observed Mahalanobis distance can be decomposed as follows:
\begin{eqnarray} \nonumber
    d_i(\widehat{\btheta}) &=& (\y_i - \textbf{X}_i\widehat{\bbeta}-\widehat{c}\,\textbf{Z}_i  \widehat{\bDelta})^\top \widehat{\bPsi}_i^{-1} (\y_i - \textbf{X}_i\widehat{\bbeta}-\widehat{c}\,\textbf{Z}_i  \widehat{\bDelta}) \\ \label{eq:MDdecomp}
    &=& \textbf{e}_i^\top \widehat{\bSigma}_i^{-1}\textbf{e}_i + \left(\widehat{\bmu}_{bi} - \widehat{c} \widehat{\bDelta} \right)^\top
    \widehat{\textbf{D}}^{-1} \left(\widehat{\bmu}_{bi} - \widehat{c} \widehat{\bDelta} \right) = d_{\textbf{e}_i}(\widehat{\btheta}) + d_{\bi}(\widehat{\btheta}),
\end{eqnarray}
where $\textbf{e}_i = \y_i-\textbf{X}_i\widehat{\bbeta}-\textbf{Z}_i \widehat{\bmu}_{bi}$ and $\widehat{\bmu}_{bi}$ is as in \eqref{eq:estbi}. This decomposition gives some insight on how the estimated random effects $\widehat{\bi}$ and the estimated residuals $\textbf{e}_i$ affect the overall distance.

Another important assumption that should be taken into account is the dependence structure assumed to the within-subject errors. In the context of time series data, a commonly used tool for investigation serial correlation is the empirical autocorrelation function (ACF) \citep{box1976time}. 
In the context of mixed models, \cite{PinheiroBates2000} proposes to use the empirical autocorrelation function for the residuals of a fitted LMM.
Based on this approach and restricting to the case that the data is observed at discrete times, %and considering that the data is sorted , 
let $\textbf{r}_i = \widehat{\bUpsilon}_i^{-1/2}\left(\y_i - \textbf{X}_i\widehat{\bbeta} \right)$ be the standardized marginal residual vector for subject $i$, where $\bUpsilon_i$ is given in \eqref{eq:varY} and $\textbf{r}_i^\top = (r_{it_1},\hdots,r_{it_{n_i}})$. The empirical autocorrelation at lag $l$ can be defined as
\begin{equation}\label{eq:ACF}
    \widehat{\rho}(l) = \frac{\sum_{i=1}^n \sum_{\{ (j,k) |t_k-t_j=l\}} r_{it_j} r_{it_k} / N(l)}{
    \sum_{i=1}^n \sum_{j=1}^{n_i} r_{it_j}^2/N(0) },
\end{equation}
where $N(\cdot)$ is the number of pairs used in the respective numerator summation. If the within-subject dependence structure is correct, $\widehat{\rho}(l)$ is expected to be close to zero. 

Since $\textbf{r}_i$'s distribution is not symmetrical, the interval estimates of ${\rho}(\cdot)$ that are commonly used in time series models are not appropriate. Alternatively, we consider a Monte Carlo estimate for a conditionally independent model, by generating $M$ samples from a CI-SMSN-LMM similar to the fitted model, calculating the standardized marginal residuals and $\widehat{\rho}(l)$ for each sample, and using empirical $100(\alpha/2)$th and $100(1-\alpha/2)$th percentiles as interval estimates of level $1-\alpha$. If the considered dependence structure is appropriate, we would expect approximately $(1-\alpha)\%$ of the empirical autocorrelations to belong to the conditionally independent interval.
}

\section{Simulation studies}\label{sec:simstudies}
In the interest of investigating empirical properties of the proposed model four simulation studies were performed, and their results are presented in this section.
In all simulation studies we initialized $\widehat{\bnu}$ as follows: $10$ for the ST distribution, $5$ for the SSL distribution, and $(0.05,0.8)$ for the SCN distribution.
Besides, for AR($p$) dependence $\bphi$ was initialized as its estimate from fitting an AR($p$)-LMM using \emph{lme} function from \emph{nlme} package in \textsf{R} \citep{nlmepackage}, while for DEC dependence  $\bphi$ was initialized by finding the maximum marginal log-likelihood function as in \eqref{eq:marginalpdfY} on a grid of $\bphi$ and for other parameters fixed.
Finally, $\bbeta, \sigma_e^2, \mathbf{D}$ and $\blambda$ were initialized at the true value plus a small random error.
For all studies, the bias and relative bias for estimating a parameter $\theta$ based on the $k$th sample is calculated by 
$\widehat{\theta}_{(k)} - \theta$ and $(\widehat{\theta}_{(k)} - \theta)/{\theta}$, respectively. The computational
procedures were implemented using the  \emph{R} software, through the
package \emph{skewlmm} \citep{skewlmm-manual}.

\subsection{First study}%sim2r3
This simulation study aims to investigate asymptotic properties of the proposed model.
Thence, we generated and estimated 500 Monte Carlo samples from the model
$$\Y_{i} = (\beta_0+b_i) \mathbf{1}_{10}+ \beta_1\x_{i}+\bepsilon_{i}, \,\, i=1,\hdots , n,$$ 
where $\beta_0=1$, $\beta_1=2$, $\mathbf{1}_{k}$ is the all-ones vector of length $k$ and $\x_i = (x_{i1},\hdots,x_{i10})^\top$, with $x_{ij}$ being generated from the $U(0,2)$ distribution, $i=1,\hdots,n$ and $j=1,\hdots,10$, with $n$ taking values $50, 100, 200$ and $350$. Let $\textbf{R}_i$ be a $10\times 10$ AR($2$) dependence matrix, as given in Subsection \ref{subsec:covstructs}, with $\phi_1=0.6$ and $\phi_2=-0.2$. Four scenarios were considered:
\begin{itemize}[\itemsep=0em ]
\item[a)] $b_i\iid \SN_1(-1.0705,2,3)$ and $\bepsilon_i\ind
\textrm{N}_{10}(\mathbf{0},0.25 \textbf{R}_i)$;
\item[b)] $b_i\iid \ST_1(-1.2324,2,3;6)$ and $\bepsilon_i\ind
\textrm{t}_{10}(\mathbf{0},0.25 \textbf{R}_i;6)$;
\item[c)] $b_i\iid \SSL_1(-1.5165,2,3;1.7)$ and $\bepsilon_i\ind
\textrm{SL}_{10}(\mathbf{0},0.25 \textbf{R}_i;1.7)$; and
\item[d)] $b_i\iid \SCN_1(-1.2915,2,3;0.25,0.3)$ and $\bepsilon_i\ind
\textrm{CN}_{10}(\mathbf{0},0.25 \textbf{R}_i;0.25,0.3)$.
\end{itemize}

The ML estimates and their associated standard errors were recorded. 
In order to examine the consistency of the approximated method to get standard errors described in Subsection \ref{sec:stderror}, we computed the standard deviation of the ML estimates obtained from the 500 Monte Carlo samples (denoted by MC-SD) and compared it with the average of the standard error estimates obtained as described in Subsection \ref{sec:stderror} (denoted by ML-SE), for each scenario.
Likewise, the average of the ML estimates will be denoted by MC-AV.

Table \ref{tab:sim2} presents results for $n=100$ and $n=350$.
In general, the estimation method of the standard errors provide results close
to the empirical ones, and the closeness improves as the number of subjects increases.
However, the standard error approximation for the skewness parameter seems to be poor.
Additionally, there is a bias in the estimates of $\mathbf{D}^{1/2}$, $\lambda$ and $\bnu$, but it gets smaller as $n$ increases.

\begin{table}[ht]\setlength{\tabcolsep}{3pt} 
\centering 
\small
\caption{Simulation study 1. Results based on 500 Monte Carlo samples with different number of subjects ($n$). MC-AV and MC-SD refer to the mean and standard deviation of the estimates, respectively. ML-SE denotes the average of standard errors obtained as described in Subsection \ref{sec:stderror} and the numbers between parenthesis are the parameter values.}
\label{tab:sim2}
\begin{tabular}{@{}c>{\columncolor[gray]{0.9}}r>{\columncolor[gray]{0.9}}r>{\columncolor[gray]{0.9}}rrrr>{\columncolor[gray]{0.9}}r>{\columncolor[gray]{0.9}}r>{\columncolor[gray]{0.9}}rrrr@{}}
\toprule
 & \multicolumn{3}{>{\columncolor[gray]{0.9}}c}{SN} & \multicolumn{3}{c}{ST} & \multicolumn{3}{>{\columncolor[gray]{0.9}}c}{SSL} & \multicolumn{3}{c}{SCN} \\\midrule
 & \multicolumn{1}{>{\columncolor[gray]{0.9}}c}{\footnotesize MC-AV} & \multicolumn{1}{>{\columncolor[gray]{0.9}}c}{\footnotesize ML-SE} & \multicolumn{1}{>{\columncolor[gray]{0.9}}c}{\footnotesize MC-SD} & \multicolumn{1}{c}{\footnotesize MC-AV} & \multicolumn{1}{c}{\footnotesize ML-SE} & \multicolumn{1}{c}{\footnotesize MC-SD} & \multicolumn{1}{>{\columncolor[gray]{0.9}}c}{\footnotesize MC-AV} & \multicolumn{1}{>{\columncolor[gray]{0.9}}c}{\footnotesize ML-SE} & \multicolumn{1}{>{\columncolor[gray]{0.9}}c}{\footnotesize MC-SD} & \multicolumn{1}{c}{\footnotesize MC-AV} & \multicolumn{1}{c}{\footnotesize ML-SE} & \multicolumn{1}{c}{\footnotesize MC-SD} \\ \midrule
 & \multicolumn{12}{c}{$n=100$} \\ \midrule
$\beta_0 \footnotesize\,(1)$ & 1.003 & 0.101 & 0.085 & 1.002 & 0.124 & 0.099 & 0.996 & 0.147 & 0.118 & 1.006 & 0.124 & 0.109 \\
$\beta_1 \footnotesize\,(2)$ & 2.002 & 0.026 & 0.024 & 1.999 & 0.027 & 0.026 & 2.003 & 0.033 & 0.032 & 1.999 & 0.029 & 0.027 \\
$\sigma_e^2 \footnotesize\,(0.25)$ & 0.250 & 0.013 & 0.012 & 0.251 & 0.020 & 0.020 & 0.253 & 0.018 & 0.025 & 0.248 & 0.016 & 0.019 \\
$\phi_1 \footnotesize\,(0.6)$ & 0.597 & 0.039 & 0.038 & 0.599 & 0.040 & 0.038 & 0.596 & 0.040 & 0.039 & 0.599 & 0.040 & 0.039 \\
$\phi_2 \footnotesize\,(-0.2)$ & -0.198 & 0.041 & 0.040 & -0.198 & 0.043 & 0.042 & -0.198 & 0.042 & 0.040 & -0.200 & 0.042 & 0.040 \\
D$^{1/2} \footnotesize\,(\sqrt{2})$ & 1.387 & 0.184 & 0.154 & 1.409 & 0.181 & 0.145 & 1.400 & 0.184 & 0.161 & 1.384 & 0.180 & 0.155 \\
$\lambda \footnotesize\,(3)$ & 3.301 & 2.318 & 1.411 & 3.371 & 2.248 & 1.391 & 3.162 & 1.965 & 1.169 & 3.272 & 2.133 & 1.342 \\
$\nu_1$ &  &  &  & 6.476 &  &  & 1.789 &  &  & 0.254 &  &  \\
$\nu_2$ &  &  &  &  &  &  &  &  &  & 0.302 &  &  \\ \midrule
 & \multicolumn{12}{c}{$n=350$} \\ \midrule
$\beta_0 \footnotesize\,(1)$ & 1.006 & 0.053 & 0.046 & 1.009 & 0.065 & 0.046 & 1.002 & 0.078 & 0.054 & 1.010 & 0.065 & 0.048 \\
$\beta_1 \footnotesize\,(2)$ & 2.000 & 0.013 & 0.013 & 1.999 & 0.014 & 0.014 & 1.999 & 0.017 & 0.017 & 1.999 & 0.015 & 0.015 \\
$\sigma_e^2 \footnotesize\,(0.25)$ & 0.250 & 0.007 & 0.006 & 0.251 & 0.011 & 0.010 & 0.252 & 0.009 & 0.013 & 0.251 & 0.008 & 0.009 \\
$\phi_1 \footnotesize\,(0.6)$  & 0.599 & 0.020 & 0.020 & 0.599 & 0.021 & 0.021 & 0.599 & 0.021 & 0.020 & 0.600 & 0.021 & 0.021 \\
$\phi_2 \footnotesize\,(-0.2)$  & -0.199 & 0.021 & 0.020 & -0.200 & 0.022 & 0.022 & -0.198 & 0.022 & 0.021 & -0.201 & 0.022 & 0.022 \\
D$^{1/2} \footnotesize\,(\sqrt{2})$ & 1.408 & 0.092 & 0.077 & 1.418 & 0.094 & 0.073 & 1.421 & 0.094 & 0.076 & 1.415 & 0.092 & 0.072 \\
$\lambda \footnotesize\,(3)$  & 3.022 & 0.796 & 0.528 & 3.056 & 0.766 & 0.453 & 3.030 & 0.771 & 0.468 & 3.020 & 0.762 & 0.446 \\
$\nu_1$ &  &  &  & 6.038 &  &  & 1.726 &  &  & 0.251 &  &  \\
$\nu_2$ &  &  &  &  &  &  &  &  &  & 0.301 &  &  \\ \bottomrule
\end{tabular}
\end{table}

\begin{figure}[ht]
    \centering
    \includegraphics[width=.95\textwidth]{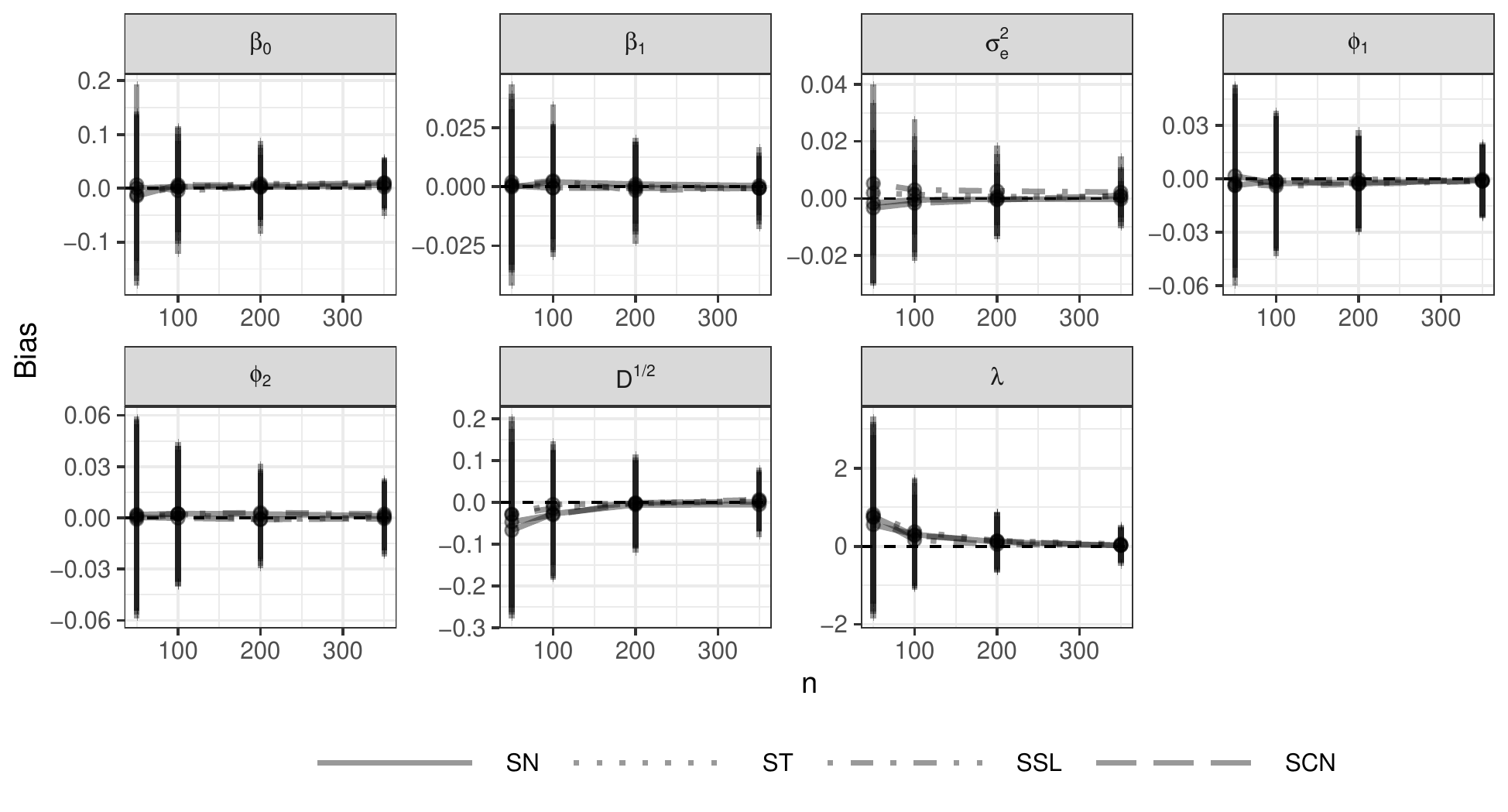}
    \caption{Simulation study 1. Mean bias and $\pm 1$ standard deviation of estimates, according to distribution and parameter, by number of subjects ($n$), based on 500 Monte Carlo samples.}
    \label{fig:sim2a}
\end{figure}

\begin{figure}[ht]
    \centering
    \includegraphics[width=.9\textwidth]{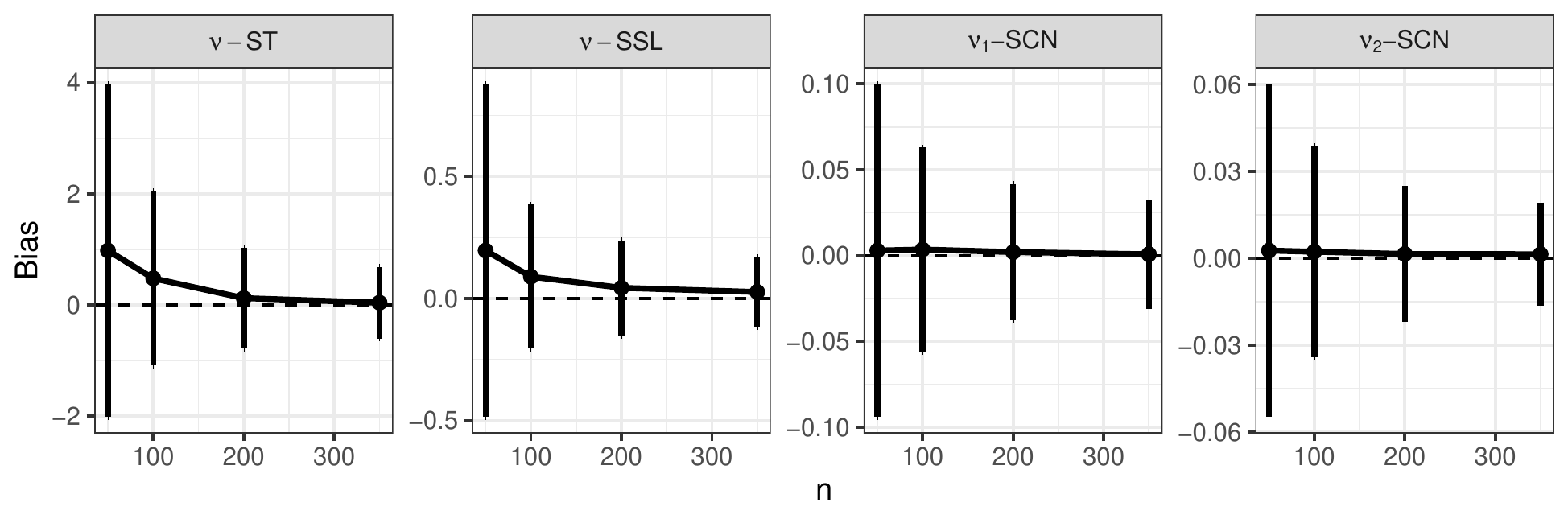}
    \caption{Simulation study 1. Mean bias and $\pm 1$ standard deviation for estimates of $\bnu$, according to distribution, by number of subjects ($n$), based on 500 Monte Carlo samples.}
    \label{fig:sim2b}
\end{figure}

The bias trend can be seem more clearly in Figures \ref{fig:sim2a} and \ref{fig:sim2b}, which present the mean bias and $\pm1$ standard deviation for each distribution and parameter, by number of subjects.
As the number of subjects increases, the bias (when it exists) gets closer to zero and its standard deviation gets smaller, indicating consistency of the estimators. 

\subsection{Second study}\label{subsec:simstudy1} %sim2r1 (st) and sim2r2 (sn) -> sim6

In order to evaluate the performance of the proposed model and the impact in estimating with the wrong distribution, we generated 500 Monte Carlo data sets from the following model:
$$\Y_{i} = (\beta_0+b_{0i}) \mathbf{1}_{10} + (\beta_1+ b_{1i})\x_{1i}+\beta_2\x_{2i}+\bepsilon_{i}, \,\, i=1,\hdots , 100,$$ 
where $\beta_0=1$, $\beta_1=2$, $\beta_2=1.5$, $\mathbf{1}_{k}$ is the all-ones vector of length $k$, $\x_{1i} = (1,2,\hdots,10)^\top$ and $\x_{2i} = w_i \mathbf{1}_{10}$ is a group indicator taken as $w_i=0$ if $i\leq 50$ and $w_i=1$ if $i> 50$. Let $\textbf{R}_i$ be the AR($2$) dependence matrix, as given in Subsection \ref{subsec:covstructs}, with $\phi_1=0.6$ and $\phi_2=-0.2$ and let $\be_i = (b_{0i},b_{1i})^\top$. For data generation, two scenarios were considered:
\begin{itemize}[\itemsep=0em ] 
    \item[a)] $\be_i\iid \SN_2 \left(
    \left[\begin{array}{c}
         -0.4718 \\ -0.7500
    \end{array} \right] ,
    \left[\begin{array}{cc}
         2.0& 0.2\\ 0.2& 1.0
    \end{array} \right] ,
    \left[\begin{array}{c}
         2 \\ 5
    \end{array} \right] \right)$ and $\bepsilon_i\ind
\textrm{N}_{10}(\mathbf{0},0.25 \textbf{R}_i)$; and
\item[b)] $\be_i\iid \ST_2\left(
    \left[\begin{array}{c}
         -0.5432 \\ -0.8635
    \end{array} \right] ,
    \left[\begin{array}{cc}
         2.0& 0.2\\ 0.2& 1.0
    \end{array} \right] ,
    \left[\begin{array}{c}
         2 \\ 5
    \end{array} \right] ;6\right)$ and $\bepsilon_i\ind
\textrm{t}_{10}(\mathbf{0},0.25 \textbf{R}_i;6)$. 
\end{itemize}

\begin{figure}[ht]
    \centering
    \includegraphics[width=0.98\textwidth]{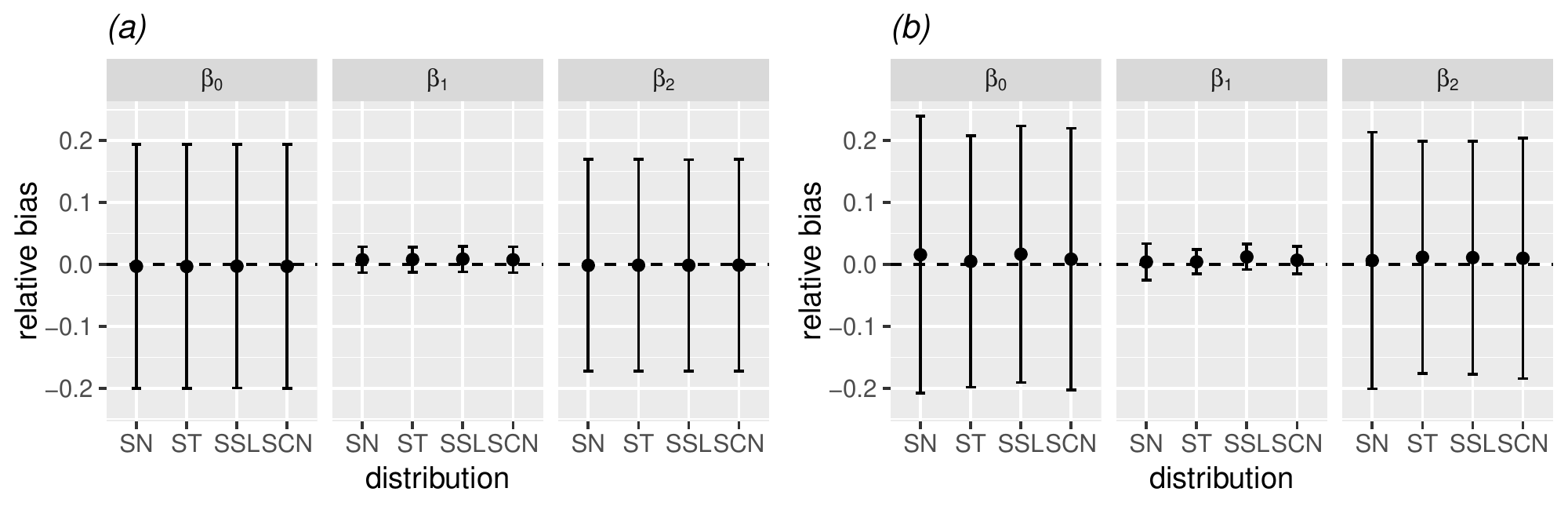}
    \caption{Simulation study 2.  Mean relative bias and $\pm 1$ standard deviation for estimates of $\beta_0$, $\beta_1$ and $\beta_2$ when generating data from both scenarios considered --{\it(a)} SN and {\it(b)} ST-- and estimating the four models.} %Boxplots of the bias of $\widehat{\beta}_0$ and $\widehat{\beta}_1$ when generating data from both scenarios considered --{\it(a)} ST and {\it(b)} SN-- and estimating the four models.}
    \label{fig:sim1}
\end{figure}

Figure \ref{fig:sim1} presents the mean relative bias of $\widehat{\beta}_0$, $\widehat{\beta}_1$ and $\widehat{\beta}_2$ for both scenarios of data generation when estimating with all the distributions considered in this paper.
When the generating model is SN, all distributions seem to fit the data equally well.
On the other hand, when the generating model is ST, estimating with wrong distributions seems to slightly increase variance of the estimator.

Furthermore, to evaluate the capability of the proposed selection criteria in selecting the appropriate distribution, we computed the AIC and BIC for each model and for each sample. Table \ref{tab:sim1} presents the number of times that each model was selected for both scenarios considered. One can see that in general the criteria can classify the correct model well, and they seem to be specially good at distinguishing between the skew-normal distribution and the heavier-tailed distributions.

\begin{table}[ht]
\centering
\caption{Simulation study 2. Number of times that each distribution was selected based on each 
        selection criterion for both scenarios considered.}
\label{tab:sim1}
\begin{tabular}{@{}c|rrrr|rrrr@{}}
\toprule
\multirow{2}{*}{Criterion} & \multicolumn{4}{c|}{Scenario (a) - SN} & \multicolumn{4}{c}{Scenario (b) - ST}\\ \cmidrule(l){2-9} & \multicolumn{1}{c}{SN} & \multicolumn{1}{c}{ST} & \multicolumn{1}{c}{SSL} & \multicolumn{1}{c|}{SCN} & \multicolumn{1}{c}{SN} & \multicolumn{1}{c}{ST} & \multicolumn{1}{c}{SSL} & \multicolumn{1}{c}{SCN} \\ \midrule
AIC& 476 &15&	6&	3&0	&373	&70	&57\\
BIC& 499&	1&	0&	0& 0	&411	&79	&10\\ \bottomrule
\end{tabular}
\end{table}

\subsection{Third study}%sim3
In the interest of analyzing the impact of specifying the wrong dependence structure on parameter estimates, we generated $500$ Monte Carlo samples from the following LMM:
$$\Y_{i} = (\beta_0+b_i) \mathbf{1}_{n_j}+ \beta_1\x_{i}+\bepsilon_{i}, \,\, i=1,\hdots , 100,$$ 
where $\beta_0=1$, $\beta_1=2$, $\mathbf{1}_{k}$ is the all-ones vector of length $k$ and $\x_i = (x_{i1},\hdots,x_{in_j})^\top$, with $x_{ij}$ being generated from the $U(0,2)$ distribution, for $i=1,\hdots,100$, $j=1,\hdots,n_j$, and $n_1=\hdots=n_{100}$ taking values $5, 10$ and $15$. Let $\textbf{R}_i$ be an $n_j \times n_j$ AR($2$) dependence matrix, as given in Subsection \ref{subsec:covstructs}, with $\phi_1=0.6$ and $\phi_2=-0.2$. For this study, we considered $b_i\iid \ST_1(-1.2324,2,3;6)$ and $\bepsilon_i\ind \textrm{t}_{n_j}(\mathbf{0},0.25 \textbf{R}_i;6)$.
Then, we estimated the ST-LMM by considering four covariance structures: CI, AR($1$), AR($2$) and DEC.

\begin{figure}[ht]
    \centering
    \includegraphics[width=.95\textwidth]{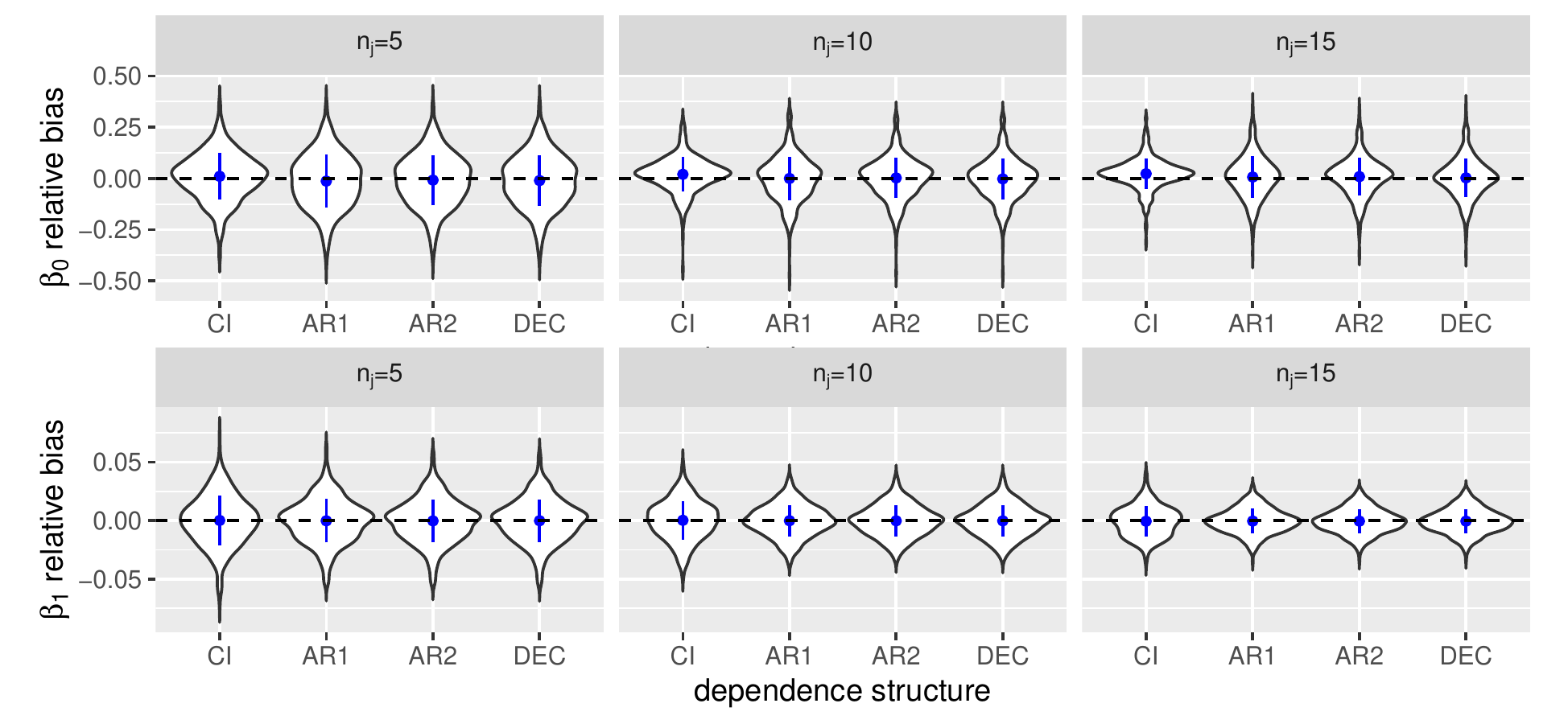}
    \caption{Simulation study 3. Violin plot of the relative bias of $\widehat{\beta}_0$ and $\widehat{\beta}_1$ when generating data from an AR($2$)-ST-SMSN and fitting a ST-LMM considering different covariance structures. The blue dot indicates the mean and the blue line indicates $\pm 1$ standard deviation.}
    \label{fig:sim3}
\end{figure}

Figure \ref{fig:sim3} presents violin plots of the relative bias of $\widehat{\beta}_0$ and $\widehat{\beta}_1$ for all dependence structures considered, showing a rotated kernel density plot and some summary statistics of the relative bias. When the number of observations per subject is small ($n_j=5$), 
the estimates are similar for all covariance structures.
As the number of observations per subject increases, the impact in considering the conditionally independent model becomes more evident,  although for the other dependence structures considered, the wrong choice of the covariance function does not seem to cause much effect on $\widehat{\bbeta}$.
For the CI model, the relative bias of $\widehat{\beta}_0$ has high density above 0, indicating a bias on the intercept estimate that increases with $n_j$, and the relative bias of  $\widehat{\beta}_1$ presents more variability for all sample sizes.

Table \ref{tab:sim3} presents the number of times that each model was selected based on AIC and BIC criteria. For $n_j=5$, the BIC criterion --which gives more penalty for the number of parameters-- select the model with AR($1$) dependence most of the times, but it does not select the CI model any time. The AIC criterion, by the other hand, select the correct model most of times for all sample sizes.
Additionally, as $n_j$ increases the selection criteria can distinguish better between the covariance structures, as expected. 

\begin{table}[ht]
\centering
\caption{Simulation study 3. Number of times that each model was selected based on each selection criterion.}
\label{tab:sim3}
\begin{tabular}{@{}r|rrrr|rrrr@{}}
\toprule
\multicolumn{1}{c|}{\multirow{2}{*}{$n_j$}} & \multicolumn{4}{c|}{AIC} & \multicolumn{4}{c}{BIC} \\ \cmidrule(l){2-9} 
 & \multicolumn{1}{c}{CI} &\multicolumn{1}{c}{AR(1)} & \multicolumn{1}{c}{AR(2)} & \multicolumn{1}{c|}{DEC} &\multicolumn{1}{c}{CI} & \multicolumn{1}{c}{AR(1)} & \multicolumn{1}{c}{AR(2)} & \multicolumn{1}{c}{DEC} \\ \midrule
 5 & 0 & 68 & 294 & 138 & 0 & 254 & 175 & 71 \\
 10 & 0 & 1 & 371 & 128 & 0 & 11 & 367 & 122 \\
 15 & 0 & 0 & 394 & 106 & 0 & 0 & 394 & 106\\ \bottomrule
\end{tabular}
\end{table}

Once the parameter estimates of $\bphi$ and $\sigma_e^2$ are not directly comparable between different covariance structures, we present in Table \ref{tab:sim3-2} the first row of the estimated within-subject variance matrix based on the average of parameter estimates obtained for all dependence structures.
	
\begin{table}[ht]
	\centering \small
	\caption{Simulation study 3. First row of the estimated within-subject variance matrix, based on average estimate from 500 Monte Carlo samples, for different dependence structures.} \label{tab:sim3-2}
	\begin{tabular}{@{}crrrrrrrrrr@{}}
		\toprule
		lag & \multicolumn{1}{c}{0} & \multicolumn{1}{c}{1} & \multicolumn{1}{c}{2} & \multicolumn{1}{c}{3} & \multicolumn{1}{c}{4} & \multicolumn{1}{c}{5} & \multicolumn{1}{c}{6} & \multicolumn{1}{c}{7} & \multicolumn{1}{c}{8} & \multicolumn{1}{c}{9} \\ \midrule
		True & 0.347 & 0.174 & 0.035 & -0.014 & -0.015 & -0.006 & -0.001 & 0.001 & 0.001 & 0.000 \\
		CI & 0.291 & 0.000 & 0.000 & 0.000 & 0.000 & 0.000 & 0.000 & 0.000 & 0.000 & 0.000 \\
		AR(1) & 0.393 & 0.220 & 0.124 & 0.069 & 0.039 & 0.022 & 0.012 & 0.007 & 0.004 & 0.002 \\
		AR(2) & 0.348 & 0.174 & 0.035 & -0.013 & -0.015 & -0.006 & -0.001 & 0.001 & 0.001 & 0.000 \\
		DEC & 0.362 & 0.189 & 0.053 & 0.010 & 0.001 & 0.000 & 0.000 & 0.000 & 0.000 & 0.000 \\ \bottomrule
	\end{tabular}
\end{table}

\section{Application: Schizophrenia data}\label{sec:aplic}
Schizophrenia is a severe psychiatric disorder characterized by delusions, hallucinations, persistent delusions and sometimes disorganized behavior and speech. In order to study the equivalency of a new antipsychotic drug for schizophrenia, \cite{lapierre1990controlled} presented a double-blinded clinical trial with randomization among four treatments: three doses (low, medium and high) of a new therapy (NT) against a standard therapy (ST), for 245 patients with acute schizophrenia. 
Initial studies prior to this double-blinded study suggested that the experimental drug had equivalent antipsychotic activity, with less side effects. 
The study was conducted at 13 clinical centres, and the primary response variable was assessed using the Brief Psychiatric Rating Scale (BPRS) at baseline (week 0), and at weeks 1, 2, 3, 4 and 6 of treatment. 
This scale measures the extent of a total of 18 features, and rates each one on a seven point scale, with a higher number reflecting a worse evaluation. The total BPRS score is the sum of the scores on the 18 items.

\begin{figure}[ht]
	\centering
	\includegraphics[width=.95\textwidth]{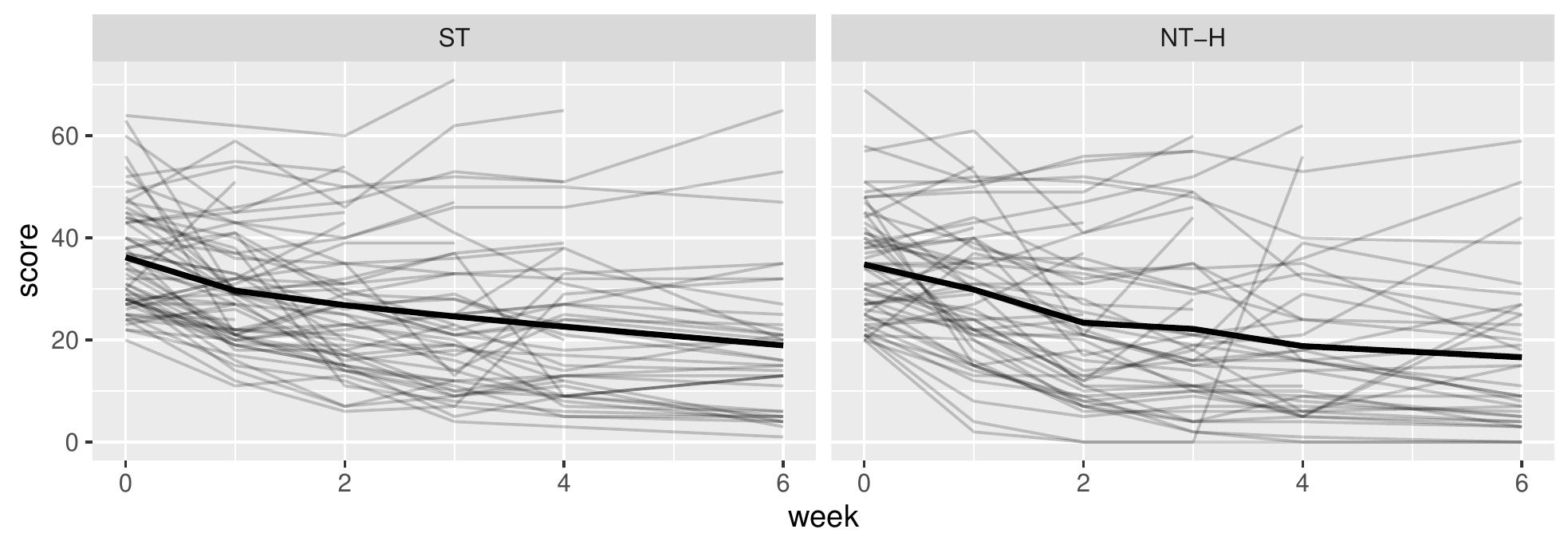}
	\caption{Trajectories of schizophrenia levels for the data. The thicker solid line indicates the mean profile in the treatment.}
	\label{schizo}
\end{figure}

Figure \ref{schizo} displays individual BPRS trajectories evolved over six visits
along with their mean profiles for the standard therapy and the high dose of the new therapy. For simplicity, we will consider only this two treatments (118 patients), but an extension for modeling the four treatments is straightforward.
\cite{ho2010robust} showed that both subject-specific intercepts and
slopes are positively skewed and that there are outliers at the level of the random effects, indicating the need of a robust model that accommodates the random effect skewness. \cite{ho2010robust} suggests a robust linear mixed model using the skew t distribution, however the paper does not take into account a possible serial correlation, despite the fact that the repeated measures of each subject were collected over time.

Thus, it is of practical interest to develop a
statistical model with considerable flexibility in the distributional assumptions of the random effects as well as the error terms, and that can accommodate some possible within-subject serial correlation.
Following \cite{ho2010robust} and based on the trajectories presented in Figure \ref{schizo}, we propose to fit the model 
$$ \Y_{i} = (\beta_0+ b_{0i})\mathbf{1}_{n_i} + (\beta_1+b_{1i}) {\bf x}_i+ \beta_2 {\bf x}_i^2 +\beta_3\,\textbf{NT}_i+ \bepsilon_{i}, \,\, i=1,\hdots , 118,$$
where $\Y_{i}$ is the BRPS score vector divided by 10 for the $i$th participant, $\mathbf{1}_{n_i}$ is the all-ones vector of length $n_i$, ${\bf x}_i=(x_{i1},\hdots,x_{in_i})^\top$, with $x_{ij}$ taken as (time - 3)/10 with time being measured in weeks from the baseline, ${\bf x}^2_i = (x_{i1}^2,\hdots,x_{in_i}^2)^\top$, and $\textbf{NT}_i$ is an all-ones vector if the $i$th subject received the new therapy and an all-zeros vector otherwise. 
We fit the SMSN-LMM considering the covariance structures presented in Subsection \ref{subsec:covstructs} and preserving the last three observations from subjects with identification (ID) numbers $204$ and $1608$ for prediction evaluation purposes.

Table \ref{tab:schizo1} presents AIC and BIC criteria for all distributions and covariance structures considered. The lowest value for both criteria is the one from the AR($1$)-ST-LMM and therefore this model is selected for further analyses. Table \ref{tab:schizo2} summarizes the results from ML estimation of the AR($1$)-ST-LMM. Moreover, 95\% confidence intervals were calculated for $\bbeta$ by considering the asymptotic normal approximation for the distribution of ML estimators.
We conclude that all parameters are significantly different from $0$, except for $\beta_3$, the estimate of NT effect. 
This result corroborates the equivalent hypothesis of the new antipsychotic drug.

\begin{table}[ht]
	\caption{Selection criteria for fitting the SMSN-LMM to the schizophrenia data set. Bold values indicate the smallest value from each criterion.}
	\label{tab:schizo1}
	\centering \small
	\begin{tabular}{@{}c|rrrrr|rrrrr@{}}
		\toprule
		& \multicolumn{5}{c|}{AIC} & \multicolumn{5}{c}{BIC} \\ \cmidrule(l){2-11} 
		\multirow{-2}{*}{distribution} & \multicolumn{1}{c}{CI} & \multicolumn{1}{c}{AR($1$)} & \multicolumn{1}{c}{AR($2$)} & \multicolumn{1}{c}{AR($3$)} & \multicolumn{1}{c|}{DEC} & \multicolumn{1}{c}{CI} & \multicolumn{1}{c}{AR($1$)} & \multicolumn{1}{c}{AR($2$)} & \multicolumn{1}{c}{AR($3$)} & \multicolumn{1}{c}{DEC} \\ \midrule
		SN & 1566.4 & 1542.1 & 1544.1 & 1542.7 & 1544.1 & 1610.2 & 1590.3 & 1596.6 & 1599.7 & 1596.6 \\
		ST & 1489.6 & \bf 1456.1 & 1458.1 & 1457.2 & 1457.9 & 1537.8 &\bf 1508.7 & 1515.1 & 1518.5 & 1514.9 \\
		SSL & 1499.0 & 1464.6 & 1466.5 & 1465.4 & 1466.3 & 1547.2 & 1517.2 & 1523.5 & 1526.7 & 1523.3 \\
		SCN & 1502.7 & 1470.2 & 1472.1 & 1470.8 & 1471.9 & 1555.3 & 1527.2 & 1533.5 & 1536.6 & 1533.3\\ \bottomrule
	\end{tabular}
\end{table}

Since simulations studies showed that the estimated standard error of the skewness parameter is not reliable, instead of presenting its estimate we performed a likelihood ratio test for testing the hypothesis $\Ha_0:\blambda=\mathbf{0}$, as described in Subsection \ref{subsec:lrt}.
{Since there are two restrictions under $\Ha_0$, the asymptotic distribution of $\Lambda_n$ is $\chi^2_2$. We have $\Lambda_n = 35.832$, hence the p-value of the likelihood ratio test is $P(\chi_2^2> 35.832)\approx 0$ and we conclude that the asymmetric model is necessary for modeling the schizophrenia data set, corroborating previous studies.}

\begin{table}[ht]
	\centering
	\caption{ML results from fitting the AR($1$)-ST-LMM to the schizophrenia data set, where $\mathbf{F} = \mathbf{D}^{1/2}$.} \label{tab:schizo2}
	\begin{tabular}{@{}crrrr@{}}
		\toprule
		Parameter & \multicolumn{1}{c}{Estimate} & \multicolumn{1}{c}{SE} & \multicolumn{2}{c}{95\% CI} \\ \midrule
		$\beta_0$ & 2.58 & 0.18 & 2.22 & 2.93 \\
		$\beta_1$ & -1.57 & 0.43 & -2.40 & -0.73 \\
		$\beta_2$ &5.97 & 0.76 & 4.48 & 7.46 \\
		$\beta_3$ & -0.17 & 0.16 & -0.47 & 0.14 \\
		$\sigma_e^2$ &  0.27 & 0.05 &  &  \\
		$\phi$ & 0.60 & 0.16 &  &  \\
		$\mathbf{F}_{11}$ &1.13 & 0.41 &  &  \\
		$\mathbf{F}_{12}$ &  1.24 & 0.30 &  &  \\
		$\mathbf{F}_{22}$ & 2.32 & 0.68 &  &  \\
		$\lambda_1$ & 10.53 & \multicolumn{1}{c}{-} &  &  \\
		$\lambda_2$ & 13.15 & \multicolumn{1}{c}{-} &  &  \\
		$\nu$ & 4.11 & \multicolumn{1}{c}{-} &  & \\ \bottomrule
	\end{tabular}
\end{table}

\begin{figure}[ht]
	\centering
	\includegraphics[width=.95\textwidth]{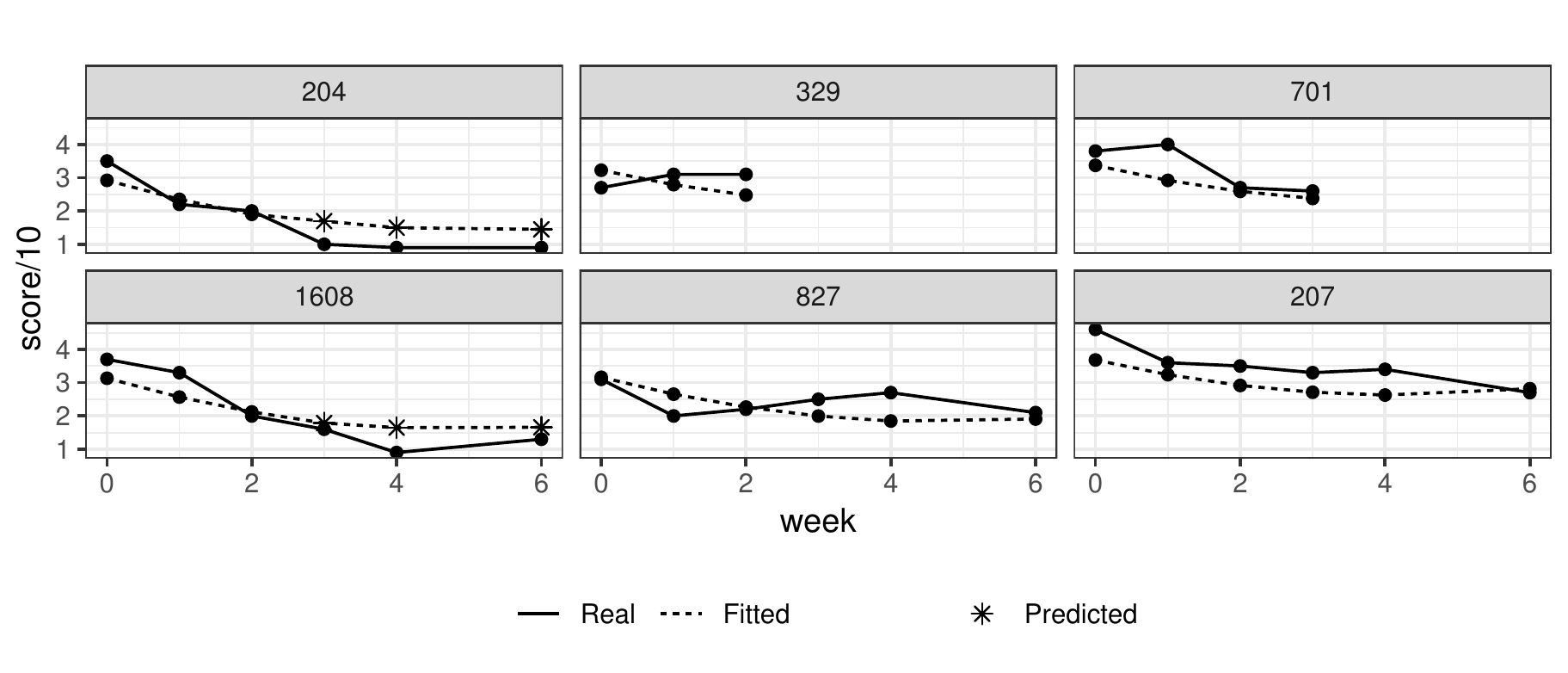}
	%\caption{(a) Evaluation of fit and prediction for six random subjects, who are identified by their ID on title, and (b) scatter plot of estimated random effects overlapped on a contour plot of the estimated marginal pdf.}
	\caption{Evaluation of fit and prediction for six random subjects of the schizophrenia data set, who are identified by their ID on title.}
	\label{fig:schizofit}
\end{figure}

\begin{figure}[ht]
	\centering
	\includegraphics[width=.7\textwidth]{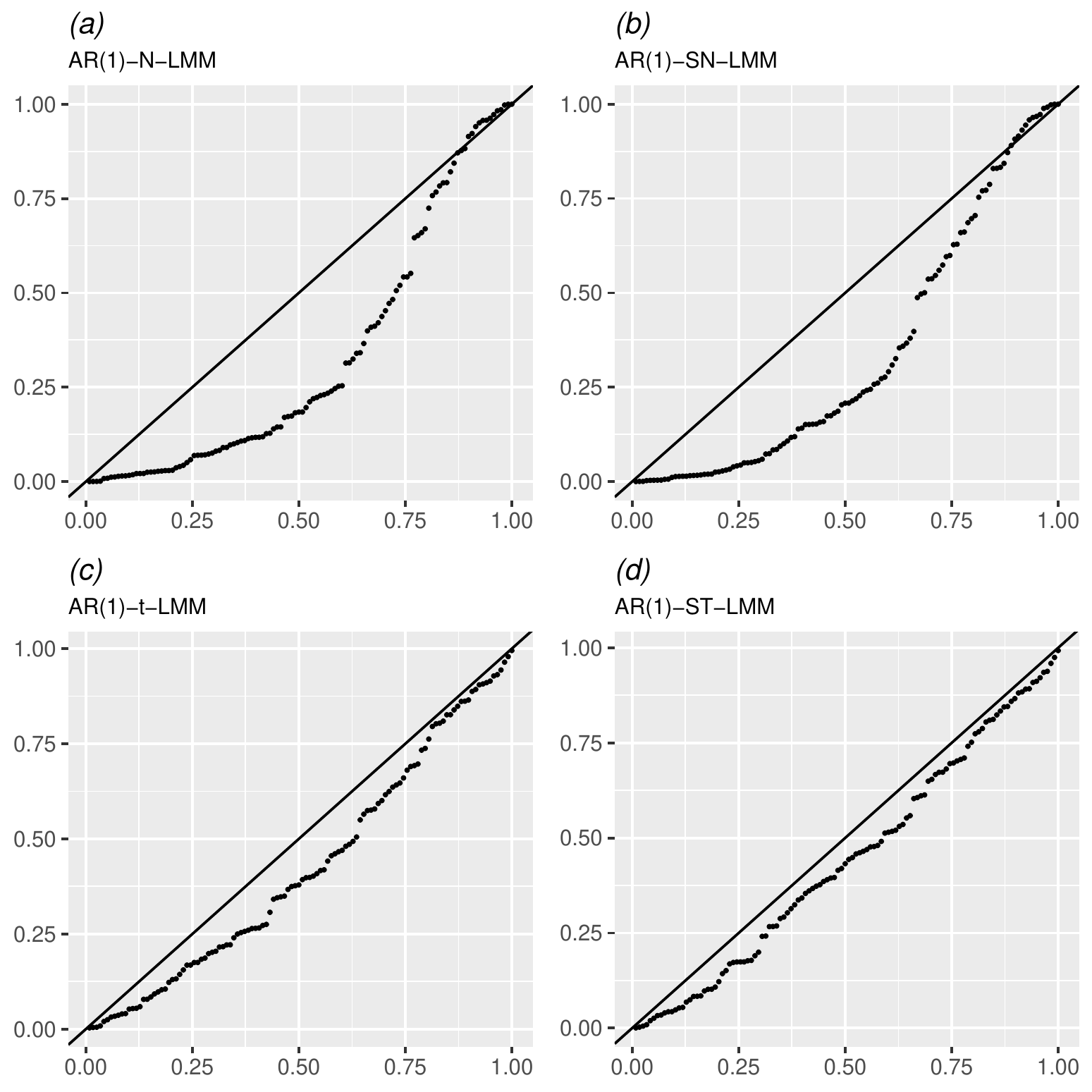}
	\caption{Healy’s plot for assessing the goodness-of-fit of fitted models.}
	\label{fig:schizofitHealey}
\end{figure}

Figure \ref{fig:schizofit} presents trajectories of six random subjects along with their fitted and predicted values, indicating adequacy of the proposed model.
Moreover, to assess the goodness of fit of the selected model, we construct a Healy-type plot, as described in Subsection \ref{subsec:residuals}, for the AR(1)-N-LMM, AR(1)-SN-LMM, AR(1)-t-LMM and AR(1)-ST-LMM. 
Since Healy's plot requires all subject to have the same number of observations, for this analysis the non-observed measurements were imputed by the prediction procedure described in Subsection \ref{subsec:prediction}. 
From Figure \ref{fig:schizofitHealey} one can see that the skew-t model \emph{(d)} is closer to the identity line than the competitive models, corroborating with the likelihood ratio test result for the asymmetry parameter and illustrating the gain in considering heavy-tailed distributions.

\begin{figure}[ht]
	\centering
	\includegraphics[width=.8\textwidth]{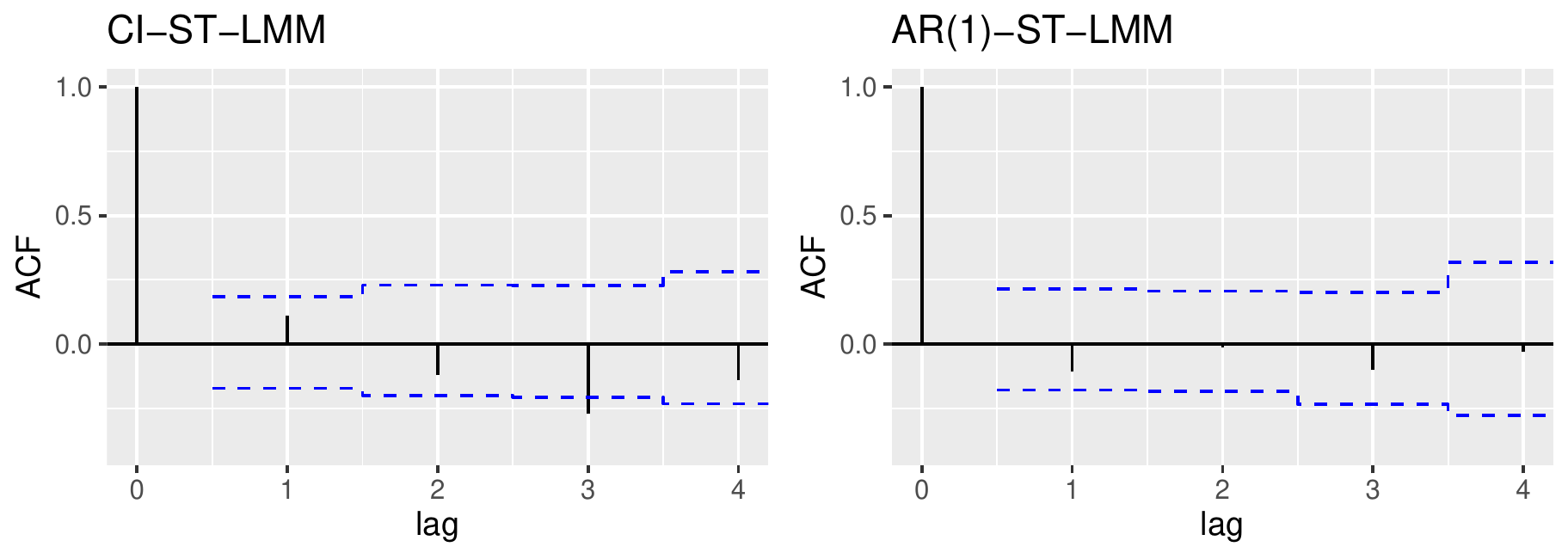}
	\caption{Autocorrelation plot based on standardized marginal residuals from CI-ST-LMM and AR($1$)-ST-LMM fitted to the schizophrenia data set. The blue dashed lines indicate a Monte Carlo $95\%$ confidence interval for a true conditionally independent model.}
	\label{fig:schizofitACF}
\end{figure}

In order to evaluate the adequacy of the selected dependence structure, we compute the empirical autocorrelations as in \eqref{eq:ACF} for CI-ST-LMM and AR(1)-ST-LMM. The results are presented in Figure \ref{fig:schizofitACF} and show that the empirical autocorrelation at lag $3$ is significantly different from $0$ for the conditionally independent model, indicating that there are some dependence not accommodated by the CI-ST-LMM, where as for the autoregressive model no significant autocorrelations are observed, indicating that the correlation structured from the AR(1)-ST-LMM is appropriate.

\begin{figure}[ht]
	\centering
	\includegraphics[width=.85\textwidth]{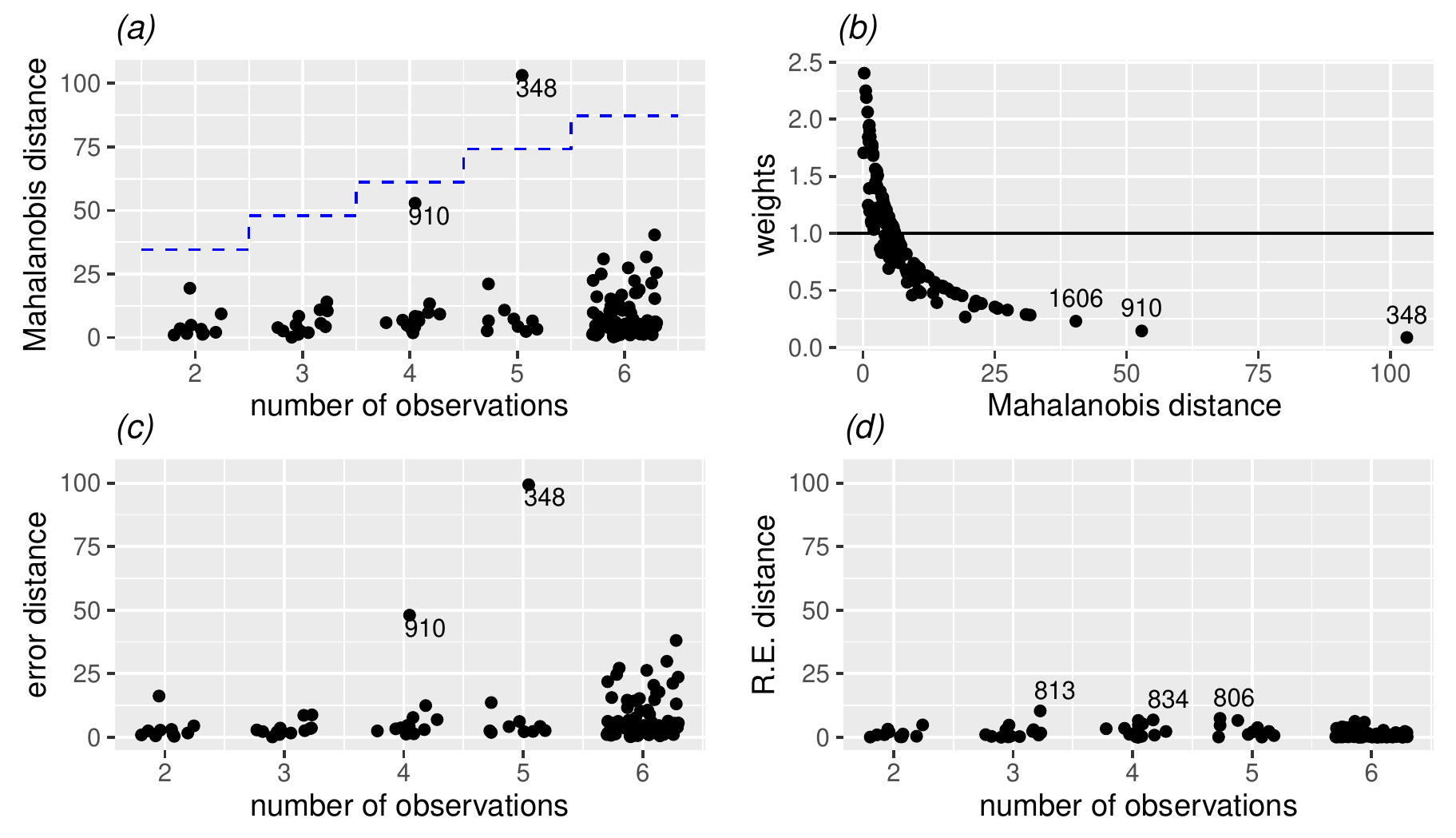}
	\caption{Mahalanobis distances \emph{(a)} and its decomposition \emph{(c)} and \emph{(d)} by number of observations for each subject and Mahalanobis distances versus $\widehat{u}_i$ \emph{(b)} for the AR($1$)-ST-LMM fitted to the schizophrenia data set. The blue dashed line indicates the theoretical $99\%$ quantile as discussed in Subsection \ref{subsec:smsn} and the black line indicates the weight for the skew-normal model.} 
	\label{fig:schizofitMD}
\end{figure}

In the interest of detecting outlying observations, Figure \ref{fig:schizofitMD}\emph{(a)} presents the Mahalanobis distance and the $99\%$ quantile from its theoretical distributions (as discussed in Subsection \ref{subsec:residuals}), by number of observations, once not all patients concluded the study.
From this analysis, only subject with ID $348$ is classified as a possible outlier. 
Furthermore, from Figure \ref{fig:schizofitMD}\emph{(b)} one can see that the weights ($\widehat{u}_i$) are close to zero for the subjects with large Mahalanobis distance, illustrating the distribution's capability of accommodating discrepant observations, in spite of the skew-normal distribution (in which all observations have the same weight).
Finally, plots \emph{(c)} and \emph{(d)} show the decomposition of the Mahalanobis distance as in 
\eqref{eq:MDdecomp}, suggesting outlying observations only at the within-subject level.

\section{Concluding remarks}\label{sec:conclude}
We proposed a likelihood approach for estimation via an EM-type algorithm, model evaluation and inference of linear mixed models under scale mixture of skew-normal distributions with within-subject correlation, considering some useful dependence structures. This work generalizes the results of \cite{Lachos_Ghosh_Arellano_2009} by developing some additional tools and making robust inferences in practical data analysis. Several simulation studies were performed in order to evaluate the proposed model.  The proposed methods were implemented as part of the new \textsf{R} package \emph{skewlmm} \citep{skewlmm-manual}, which is available for download at the CRAN repository \citep{rmanual}.

%Future works in this line include the development of diagnostic analysis in SMSN-LMM, in order to better evaluate goodness of fit and to enable detection of influential observations.
A promising avenue for future research is to consider the class of generalized hyperbolic
(GH) distributions \citep{browne2015mixture} which is generated by a variance-mean mixture of a multivariate Gaussian with a generalized
inverse Gaussian (GIG) distribution. This rich family of GH distributions include some well-known heavy-tailed and symmetric multivariate distributions including the Normal Inverse Gaussian and some members of the family of scale-mixture of skew-normal distributions.

\section*{Acknowledgments}
This study was financed in part by the Coordena\c{c}\~ao de Aperfeiçoamento de Pessoal de N\'ivel Superior - Brasil (CAPES) - Finance Code 001, and by 
Conselho Nacional de Desenvolvimento Científico e Tecnológico - Brasil (CNPq).

\newpage
\bibliographystyle{natbib}
%\bibliographystyle{elsarticle-harv}

%\bibliography{biblio}

\newpage
\appendix

\noindent{\bf Appendix }
\section{Derivatives of information matrix} \label{apsec:infmat}

Substituting $H$ in equations \eqref{eq:IPhi} and \eqref{eq:Iphi} for each distribution considered in this work, yields the following results:
\begin{itemize}
	\item[$\bullet$] {\it Skew--$t$:}
	\begin{eqnarray*}
		I^{\Phi}_i(w)&=&\frac{2^w\nu^{\nu/2}\Gamma(w+\nu/2)}{\Gamma(\nu/2)(\nu+\rm{d}_i)^{\nu/2+w}}T
		\left(\sqrt{\frac{\nu+2w}{\rm{d}_i+\nu}}\,A_i;\nu+2w\right),\,
		\, \, {\rm and}\, \,\\
		I^{\phi}_i(w)&=&\frac{2^{w}\nu^{\nu/2}\Gamma(\frac{\nu}{2}+w)}
		{\sqrt{2\pi}\Gamma(\nu/2)(\rm{d}_i+A_i^2+\nu)^{\frac{\nu}{2}+w}}.
	\end{eqnarray*}
	\item[$\bullet$]\noindent{\it Skew--slash:}
	\begin{eqnarray*}
		I^{\Phi}_i(w)&=&\frac{\nu
			2^{\nu+w}\Gamma(\nu+w)}{\rm{d}_i^{\nu+w}}P_1\left(\nu+w,\frac{\rm{d}_i}{2}\right)E\{\Phi(S_i^{1/2}A_i)\},
		\,\, \, \,  {\rm  and} \, \, \, \, \, \ \\
		I^{\phi}_i(w)&=&\frac{\nu2^{\nu+w}\Gamma(\nu+w)}{\sqrt{2\pi}(\rm{d}_i+A^2_i)^{\nu+w}}
		P_1\left(\nu+w,\frac{\rm{d}_i+A_i^2}{2}\right),
	\end{eqnarray*}
	where $S_i\sim \textrm{Gamma}(\nu+w,\frac{\rm{d}_i}{2})\mathbb{I}_{(0,1)}$.
	\item[$\bullet$]\noindent{\it Contaminated skew--normal:}
	\begin{eqnarray*}
		I^{\Phi}_i(w)&=&\sqrt{2\pi}\{\nu_1\nu_2^{w-1/2}\phi_1(\sqrt{\rm{d}_i};0,\frac{1}{\nu_2})
		\Phi(\nu_2^{1/2}A_i)+(1-\nu_1)\phi_1(\sqrt{\rm{d}_i})\Phi(A_i)\},
		\, \, \,\,
		{\rm and } \, \, \, \\
		I^{\phi}_i(w)&=&\sqrt{2\pi}\{\nu_1\nu_2^{w-1/2}\phi_1(\sqrt{\rm{d}_i};0,\frac{1}{\nu_2})
		\phi_1(\nu_2^{1/2}A_i)+(1-\nu_1)\phi_1(\sqrt{\rm{d}_i})\phi_1(A_i)\}.
	\end{eqnarray*}
\end{itemize}

In order to obtain the score vector in \eqref{eq:score}, we also need the derivatives given next.

\noindent Let $\dot{\textbf{F}}_r =\displaystyle\frac{\partial \textbf{F}}{\partial \alpha_r},\,\,r=1,\ldots,q(q + 1)/2$ and $\dot{\textbf{R}}_s =\displaystyle\frac{\partial \textbf{R}}{\partial \phi_s},\,\,s=1,\ldots,p$.

\noindent$\bullet$ {\bf For  $\log{|\bPsi_i|}$:}
\begin{eqnarray*}
	\frac{\partial \log{|\bPsi_i|}}{\partial \btau}&=&\textbf{0},\,\,\,
	\textrm{for}\,\,\, \btau=\bbeta,\blambda\,\,\, \textrm{and}\,\, \bnu,
	\\
	\frac{\partial \log{|\bPsi_i|}}{\partial
		\sigma_e^2}&=&\textrm{tr}(\bPsi^{-1}_i\textbf{R}_i),\\%\,\,
	\frac{\partial \log{|\bPsi_i|}}{\partial
		\alpha_r}&=&\textrm{tr}(\bPsi^{-1}_i
	\Z_i(\dot{\textbf{F}}_r\textbf{F}+\textbf{F}\dot{\textbf{F}}_r)\Z^{\top}_i),\,\,\, \textrm{for}\,\,\, r=1,\hdots,q(q+1)/2,\\
	\frac{\partial \log{|\bPsi_i|}}{\partial \phi_s}&=&\sigma^2_e \,\textrm{tr}(\bPsi^{-1}_i\dot{\textbf{R}}_{is}), \,\,\, \textrm{for}\,\,\, s=1,\hdots, p.	
\end{eqnarray*}

\noindent$\bullet$ {\bf For $A_i$:}
\begin{eqnarray*}
	\frac{\partial A_i}{\partial \bbeta}&=&-\frac{1} {a_i}\X^{\top}_i\bPsi_i^{-1}\Z_i\textbf{F}\blambda,\\
	\frac{\partial A_i}{\partial \blambda}&=&
	\frac{1}{a_i}\textbf{F}\Z^{\top}_i\bPsi_i^{-1}(\y_i-\X_i\bbeta-2c \Z_i\bDelta)-\frac{1}{a^2_i}A_i
	\textbf{F}^{-1}\bLambda_i \textbf{F}^{-1}\blambda + \frac{c\,d_i}{a_i ({1+\blambda^\top\blambda})}\bdelta,\\ 
	\frac{\partial A_i}{\partial
		\sigma^2_e}&=&-\frac{1}{a_i}\blambda^{\top}\textbf{F}\Z^{\top}_i\bPsi^{-1}_i
	\textbf{R}_i\bPsi^{-1}_i(\y_i-\X_i\bbeta-c \Z_i\bDelta) - \frac{1}{2\sigma^4_ea^2_i}A_i\blambda^{\top}\textbf{F}^{-1}\bLambda_i\Z^{\top}_i\textbf{R}^{-1}_i\Z_i\bLambda_i\textbf{F}^{-1}\blambda
	,\\
	\frac{\partial A_i}{\partial \alpha_r}&=&\frac{1
	}{a_i}\blambda^{\top}\dot{\textbf{F}}_r\Z^{\top}_i\bPsi_i^{-1}(\y_i-\X_i\bbeta-c \Z_i\bDelta)
	-\frac{c}{a_i}\bDelta^\top \Z_i^\top \bPsi_i^{-1}\Z_i\dot{\textbf{F}}_r\blambda 
	\\ &&-\frac{1}{a_i}\blambda^{\top}\textbf{F}\Z^{\top}_i\bPsi^{-1}_i\Z_i(\dot{\textbf{F}}_r\textbf{F}+\textbf{F}\dot{\textbf{F}}_r)\Z^{\top}_i\bPsi^{-1}_i(\y_i-\X_i\bbeta-c \Z_i\bDelta)
	\\
	&&+ \frac{1}{2a^2 _i}A_i \blambda^{\top}\textbf{F}^{-1}(
	\dot{\textbf{F}}_r\textbf{F}^{-1}\bLambda_i+\bLambda_i\textbf{F}^{-1}\dot{\textbf{F}}_r-\bLambda_i\textbf{F}^{-1}(\dot{\textbf{F}}_r\textbf{F}^{-1}+\textbf{F}^{-1}\dot{\textbf{F}}_r)\textbf{F}^{-1}\bLambda_i)\textbf{F}^{-1}\blambda, \\
	&& \textrm{for}\,\,\, r=1,\hdots,q(q+1)/2,\\
	\frac{\partial A_i}{\partial \phi_s}&=&-\frac{\sigma_e^2}{a_i}\blambda^{\top}\textbf{F}\Z^{\top}_i\bPsi^{-1}_i
	\dot{\textbf{R}}_{is}\bPsi^{-1}_i(\y_i-\X_i\bbeta-c \Z_i\bDelta)\\
	&&-\frac{1}{2\sigma^2_ea^2_i}A_i\blambda^{\top}\textbf{F}^{-1}\bLambda_i\Z^{\top}_i\textbf{R}^{-1}_i\dot{\textbf{R}}_{is}\textbf{R}^{-1}_i\Z_i\bLambda_i\textbf{F}^{-1}\blambda, \,\,\, \textrm{for}\,\,\, s=1,\hdots, p,	
\end{eqnarray*}
where $a_i=(1+\blambda^{\top}\mathbf{F}^{-1}\bLambda_i\mathbf{F}^{-1}\blambda)^{1/2}$ and $d_i =\blambda^{\top}\mathbf{F} \Z_i^\top\bPsi^{-1}_i\Z_i \mathbf{F}\blambda$.

\noindent$\bullet$ {\bf For $d_i$:}
\begin{eqnarray*}
	\frac{\partial d_i}{\partial
		\bbeta}&=&-2\X^{\top}_i\bPsi_i^{-1}(\y_i-\X_i\bbeta-c \Z_i\bDelta),\\
	\frac{\partial d_i}{\partial \blambda}&=&\frac{-2c}{\sqrt{1+\blambda\top\blambda}}\left(\mathbf{F} -\bdelta\bDelta^\top\right)\Z_i^\top\bPsi_i^{-1} (\y_i-\X_i\bbeta-c \Z_i\bDelta),\\
	\frac{\partial d_i}{\partial
		\sigma_e^2}&=&-(\y_i-\X_i\bbeta-c \Z_i\bDelta)^{\top}\bPsi^{-1}_i
	\textbf{R}_i\bPsi^{-1}_i(\y_i-\X_i\bbeta-c \Z_i\bDelta),\\
	\frac{\partial
		d_i}{\partial
		\alpha_r}&=&-(\y_i-\X_i\bbeta-c \Z_i\bDelta)^{\top} \bPsi^{-1}_i\Z_i(\dot{\textbf{F}}_r\textbf{F}+\textbf{F}\dot{\textbf{F}}_r)\Z^{\top}_i\bPsi^{-1}_i(\y_i-\X_i\bbeta-c \Z_i\bDelta),\\	
	&& - 2c\bdelta^\top \dot{\textbf{F}}_r\Z_i^\top \bPsi^{-1}_i(\y_i-\X_i\bbeta-c \Z_i\bDelta),\,\,\, \textrm{for}\,\,\, r=1,\hdots,q(q+1)/2,\\
	\frac{\partial d_i}{\partial \phi_s} &=& -\sigma^2_e(\y_i-\X_i\bbeta-c \Z_i\bDelta)^{\top}\bPsi^{-1}_i\dot{\textbf{R}}_{is}\bPsi^{-1}_i(\y_i-\X_i\bbeta-c \Z_i\bDelta),\,\,\, \textrm{for}\,\,\, s=1,\hdots, p.
\end{eqnarray*}

\section{Important results} \label{apsec:lemmas}
In this section we present some results that are useful for proving the results given in this paper.

\begin{lemma}
	Let $\A_{p\times p}$ , $\B_{p\times n}$ , $\C_{n\times n}$ , and $\mathbf{D}_{n\times p}$. If $\A^{-1}$ and $\C^{-1}$ exist, then
	$$(\A+\B\C\mathbf{D})^{-1} = \A^{-1} -\A^{-1}\B(\C^{-1}+\mathbf{D}\A^{-1}\B)^{-1}\mathbf{D}\A^{-1}.$$
\end{lemma}

\vspace{-.3cm}\noindent \emph{Proof.} This result is well-known as Woodbury matrix identity \citep{woodbury1950inverting}.

\begin{lemma}
	Let $\Y\sim \textrm{N}_n(\bmu,\bSigma)$. Then for any fixed $k$-dimensional vector $\mathbf{a}$ and any $k \times n$ matrix $\B$,
	$$
	\begin{array}{l}
	a)\, \E \left\{\Phi_k(\mathbf{a}+\B\Y\mid \beeta,\bOmega) \right\} = \Phi_k(\mathbf{a}\mid\beeta -\B\bmu,\bOmega+ \B\bSigma\B^\top); \textrm{ and }\\
	b)\, \E \left\{\phi_k(\mathbf{a}+\B\Y\mid \beeta,\bOmega) \right\} = \phi_k(\mathbf{a}\mid\beeta -\B\bmu,\bOmega+ \B\bSigma\B^\top).
	\end{array}
	$$
\end{lemma}

\vspace{-.3cm}\noindent \emph{Proof.} See \cite{ArellanoLachos2005}.

\begin{lemma}
	Let $\Y\sim \textrm{N}_p(\bmu,\bSigma)$ and $\X\sim \textrm{N}_q(\beeta,\bOmega)$. Then
	\begin{eqnarray*}
		\phi_p(\y\mid \bmu+\A \x,\bSigma) \phi_q(\x\mid \beeta,\bOmega) &=&  \phi_p(\y\mid \bmu+\A \beeta,\bSigma+\A\bOmega\A^\top) \times \\
		&&\phi_q (\x\mid \beeta+ \bLambda\A^\top\bSigma^{-1}(\y-\bmu-\A\beeta),\bLambda),
	\end{eqnarray*}
	where $\bLambda = (\bOmega^{-1}+\A^\top\bSigma^{-1}\A)^{-1}$.
\end{lemma}

\vspace{-.3cm}\noindent \emph{Proof.} See \cite{ArellanoLachos2005}.

\begin{lemma}
	Let $\Z \sim \textrm{SN}_p(\blambda)$, with zero mean and identity matrix variance, then the moments generator function of $\Z$ is
	$$M_\Z(\s) = 2 \exp \left\{\frac{1}{2}\s^\top \s \right\} \Phi_1(\bdelta^\top \s),\,\, \s \in \mathbb{R}^p.$$
\end{lemma}

\vspace{-.3cm}\noindent \emph{Proof.} See \cite{lachos2004tese}.

\begin{lemma}
	Let $V\sim \textrm{gamma}(\alpha,\beta)$, then for any $a, b \in \mathbb{R}$
	$$\E\left\{\Phi\left(a\sqrt{V}+b\right) \right\} = \textrm{P}\left\{ T\leq a\sqrt{\alpha/\beta}\right\},$$
	where $T$ is a random variable distributed as a non-central t-student with $2\alpha$ degrees of freedom and non-centrality parameter $-b$.
\end{lemma}

\vspace{-.3cm}\noindent \emph{Proof.} See \cite{azzalini2003distributions}.

\begin{lemma}
	Let $\mathbf{Y}\sim \textrm{SMSN}_p(\bmu,\bSigma,\mathbf{\blambda};H)$ and $\mathbf{X}\sim \textrm{SMN}_p(\bmu,\bSigma;H)$, then for any even function $g$, $g(\mathbf{Y}-\bmu)$ has the same distribution as $g(\mathbf{X}-\bmu)$. 
\end{lemma}

\begin{corollary}
Let $\mathbf{Y}\sim \textrm{SMSN}_p(\bmu,\bSigma,\mathbf{\blambda};H)$ and $\mathbf{X}\sim \textrm{SMN}_p(\bmu,\bSigma;H)$, then the Mahalanobis distance from the asymmetrical class  $d = \left(\mathbf{Y}-\bmu\right)^\top\bSigma^{-1}\left(\mathbf{Y}-\bmu\right)$  has the same distribution as the one from the symmetrical class $\left(\mathbf{X}-\bmu\right)^\top\bSigma^{-1}\left(\mathbf{X}-\bmu\right)$.
\end{corollary}

\vspace{-.3cm}\noindent \emph{Proof (Lemma 6 and Corollary 1).} See \cite{lachos2014multivariate}.

\section{Extra simulation study}\label{apsec:extrasim}
This simulation study aims to investigate the impacts on the estimation of not correcting the mean of $\be_i$, that is, to consider the model
\begin{equation}\label{eq:modSnmisbias}
\left( \begin{array}{c}
\textbf{b}_i \\
\bepsilon_i
\end{array}\right) \ind \textrm{SMSN}_{q+n_i}
\left( \left(\begin{array}{c}
\mathbf{0} \\
\mathbf{0}
\end{array} \right), \left(\begin{array}{cc}
\mathbf{D} & \mathbf{0} \\
\mathbf{0} & \bSigma_i
\end{array} \right), \left(\begin{array}{c}
\blambda \\
\mathbf{0}
\end{array} \right), H \right),\,\,\ii.
\end{equation}
By centering the distribution of $\be_i$ in $\textbf{0}$, we get $\E\{\be_i\} = \sqrt{\frac{2}{\pi}}\bDelta \E\left\{U_i^{-1/2}\right\}$, $\ii$, where $\E\left\{U_i^{-1/2}\right\}$ is given in Table \ref{tab:resdist}.

In order to evaluate the impacts of the formulation in \eqref{eq:modSnmisbias} in $\widehat{\bbeta}$, following the idea from Subsection \ref{subsec:simstudy1} we generated 500 Monte Carlo data sets from the model 
$$\Y_{i} = 1+ 2\xp_{i}+ b_i+\bepsilon_{i}, \,\, i=1,\hdots , 100,$$ 
where $\xp_i = (x_{i1},\hdots,x_{i10})^\top$, with $x_{ij}$ being generated from the $U(0,2)$ distribution, for $i=1,\hdots,100$ and $j=1,\hdots,10$. Let $\textbf{R}_i$ be the AR($2$) dependence matrix, as given in Subsection \ref{subsec:covstructs}, with $\phi_1=0.6$ and $\phi_2=-0.2$. Two scenarios were considered:
\begin{itemize}[\itemsep=0em ] 
	\item[a)] $b_i\iid \ST_1(0,2,3,6)$ and $\bepsilon_i\ind
	\textrm{t}_{10}(\mathbf{0},0.25 \textbf{R}_i,6)$; and
	\item[b)] $b_i\iid \SN_1(0,2,3)$ and $\bepsilon_i\ind
	\textrm{N}_{10}(\mathbf{0},0.25 \textbf{R}_i)$. 
\end{itemize}

\begin{figure}[ht]
	\centering
	\includegraphics[width=0.9\textwidth]{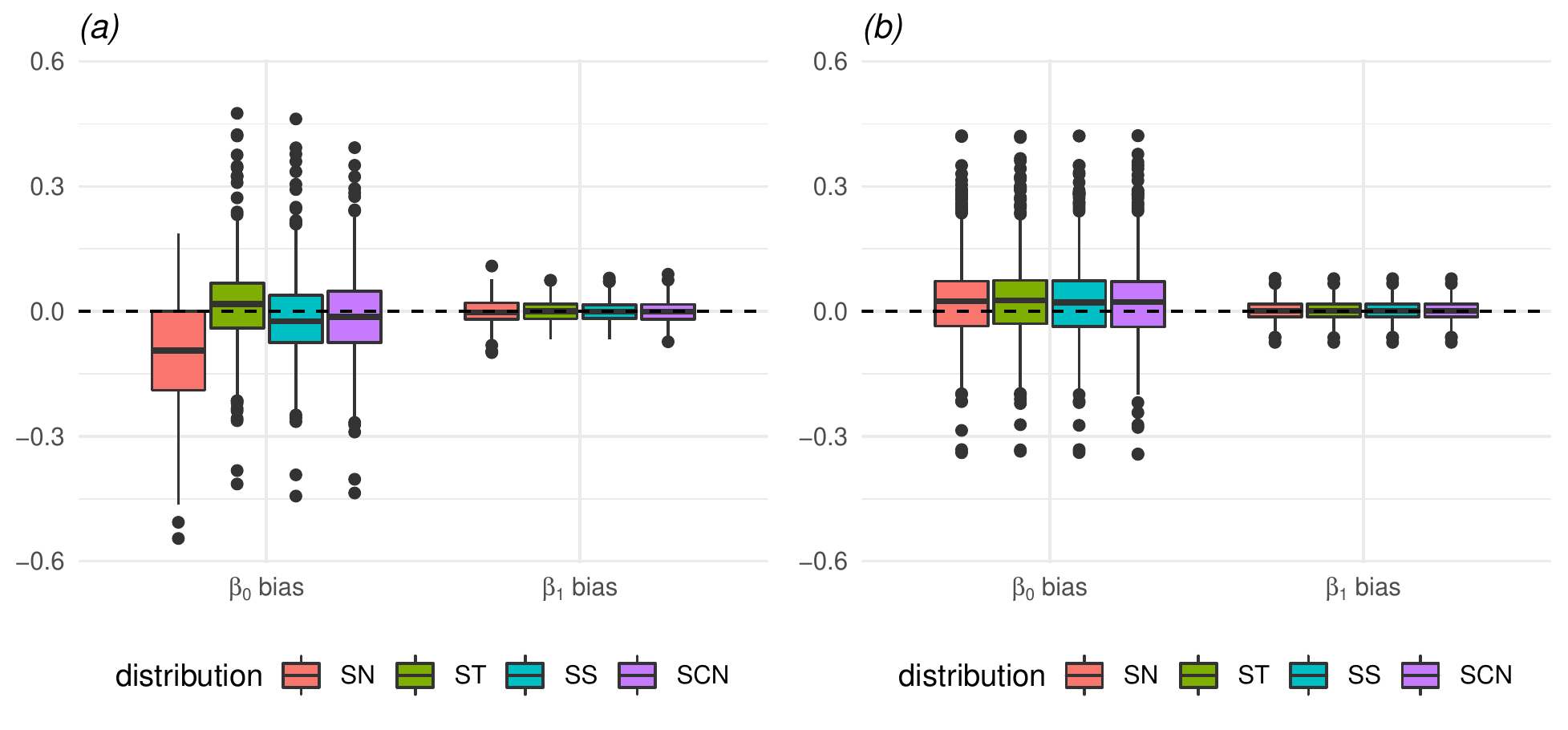}
	\caption{Extra simulation study. Boxplots of the bias of $\hat{\beta}_0$ and $\hat{\beta}_1$ when generating data from both scenarios considered --{\it(a)} ST and {\it(b)} SN-- and estimating the four models.}
	\label{fig:simextra}
\end{figure}

Figure \ref{fig:simextra} presents boxplots of the bias of $\widehat{\beta}_0$ and $\widehat{\beta}_1$ for both scenarios of data generation and for estimating the four distributions considered.
We can see a bias on the estimation of $\widehat{\beta}_0$, even when the correct distribution is considered, that does not seem to happen when the model is centered such that $\E\{\be_i\}=\mathbf{0}$, as shown in Figure \ref{fig:sim1}.
It is worth noting that the bias only appeared at the intercept, which is where the random effect was inserted.

\end{document}